\documentclass[traditabstract]{aa} 
\usepackage{graphicx}
\usepackage{txfonts}

\newcommand{\angst}{\AA}
\newcommand{\degree}{$^{\circ}$}
\newcommand{\mps}{m\,s$^{-1}${}}
\newcommand{\kw}{\mbox{$k\text{--}\omega$}}

\newcommand{\Bpar}{$B_{\parallel}$}
\newcommand{\normBpar}{\left\lvert B_{\parallel} \right\rvert}
\newcommand{\normG}{\left\lvert \Gamma_1 \right\rvert}
\newcommand\norm[1]{\left\lVert#1\right\rVert}

\usepackage{amssymb}
\usepackage{epsfig,graphicx,natbib,url}
\usepackage{color}

\begin{document}

\title{Photospheric downflows observed with SDO/HMI, \\ HINODE, and an MHD simulation}

\author {T.~Roudier\inst{1}
       \and
        M.~\v{S}vanda \inst{2,3}
       \and
       J.M.~Malherbe\inst{4}
       \and
        J.~Ballot\inst{1}
        \and
        D.~Korda \inst{2}
        \and
        Z.~Frank\inst{5}
        }

\offprints{Th. Roudier,\\
\email{thierry.roudier@irap.omp.eu}}

\institute
    {
      Institut de Recherche en Astrophysique et Plan\'etologie, Universit\'e de Toulouse, CNRS, UPS, CNES 14 avenue Edouard Belin, 31400 Toulouse, France
      \and       
      Charles University, Astronomical Institute, V Hole\v{s}ovi\v{c}k\'{a}ch 2, CZ-18000, Prague 8, Czech Republic
      \and
      Astronomical Institute of the Czech Academy of Sciences, Fri\v{c}ova 298, CZ-25165 Ondřejov, Czech Republic
     \and   
      Observatoire de Paris, LESIA, 5 place Janssen, 92195 Meudon, France, PSL Research University, CNRS, Sorbonne Universit\'es,
      Sorbonne Paris Cit\'e
      \and
      Lockheed Martin Solar and Astrophysics Laboratory, Palo Alto, 3251 Hanover Street, CA 94303, USA }

\date{Received date / Accepted date }
\titlerunning{Photospheric downflows observed with SDO/HMI, HINODE, and MHD simulation}
\authorrunning{Roudier et al.}

\abstract{

  {Downflows on the solar surface are suspected to play a major role in the dynamics of the convection zone, at least in its outer part. We investigate the existence of the long-lasting downflows whose effects influence the interior of the Sun and the outer layers.}
  {We study the sets of Dopplergrams and magnetograms observed with Solar Dynamics Observatory and Hinode spacecrafts and an magnetohydrodynamic (MHD) simulation. 
  All of the aligned sequences, which were corrected from the satellite motions and tracked with the differential rotation, were used to detect the long-lasting downflows in the quiet-Sun at the disc centre. To learn about the structure of the flows below the solar surface, the time-distance local helioseismology was used.}
  {The inspection of the 3D data cube $\left(x, y, t\right)$ of the 24-hour Doppler sequence allowed us to detect 13 persistent downflows. Their lifetimes lie in the range between 3.5 and 20 hours with sizes between $2\arcsec{}$ and $3\arcsec{}$ and speed{s} between $-0.25$ and $-0.72$~k\mps. These persistent downflows are always filled with the magnetic field with an amplitude of up to 600~Gauss. The helioseismic inversion allow us to describe the persistent downflows and compare them to the other (non-persistent) downflows in the field of view. The persistent downflows seem to penetrate much deeper and, in the case of a well-formed vortex, the vorticity keeps its integrity to the depth of about 5~Mm. In the MHD simulation, only sub-arcsecond downflows are detected with no evidence of a vortex comparable in size to observations at the surface of the Sun.}
  {The long temporal sequences from the space-borne allows us to show the existence of long-persistent downflows together with the magnetic field. They penetrate inside the Sun but are also connected with the anchoring of coronal loops in the photosphere, indicating a link between downflows and the coronal activity. A link suggests that EUV cyclones over the quiet Sun could be an effective way to heat the corona. }
}
\keywords{Sun: Granulation, Sun: Photosphere, Sun: Atmosphere}

\maketitle

\section{Introduction}

Except magnetic structures, the whole solar surface is almost completely renewed every 10--15 minutes because of the convective motions carrying the energy in the solar envelope.
At the solar surface convective cells of hot rising buoyant plasma in which energy is released by radiation exchange are observed. Around these cells, the cold plasma falls down. The cooler plasma flow is more dense than its surroundings, which  triggers the formation of turbulent plumes \citep{SN89, RZ1995} and drives the dynamics of the flow. Downdrafts are sinks where the cold plasma goes back into the Sun. These plumes  undergo secondary instabilities along their descending trajectories  producing a turbulent mixture of vorticity filaments \citep{RAST99, Bel2006, Stein2009, Rinc2018}. The vertical downflows are located in the intergranular  regions and occasionally become supersonic \citep{SN98}. Their works generally indicate the existence of strong long-lasting downdrafts penetrating into the convection zone. The models of \citet{Rast2003} predict that supergranular downflows are concentrated near the vertices of supergranular cells and merge at deeper layers. This results in large, long-lived downflow regions and, in consequence, the formation of a supergranular flow system. The kinetic energy in the convection zone is tightly connected to the evolution of the innumerable minuscule diving plumes that formed at intergranular lanes \citep{Hana2014}.

Various studies revealed the existence of vortex flows \citep{Bonet2008, Attie2009} having a wide range of spatial extents from 1 to 20~Mm with lifetimes between 5~min and 2~h. The scale of the vortex flows seem to be comparable to scales of supergranules and mesogranules. Some of the long-lasting vortex flows were located at supergranular junction vertices \citep{Attie2009, Attie2016,Req2018}. The related downflows also represent transient processes for magnetic field intensification associated with the formation of bright points in the continuum \citep{Berger2004, Bello2009, Nar2011}.

\begin{figure*}
    \centerline{\includegraphics[width=0.99\textwidth,clip=]{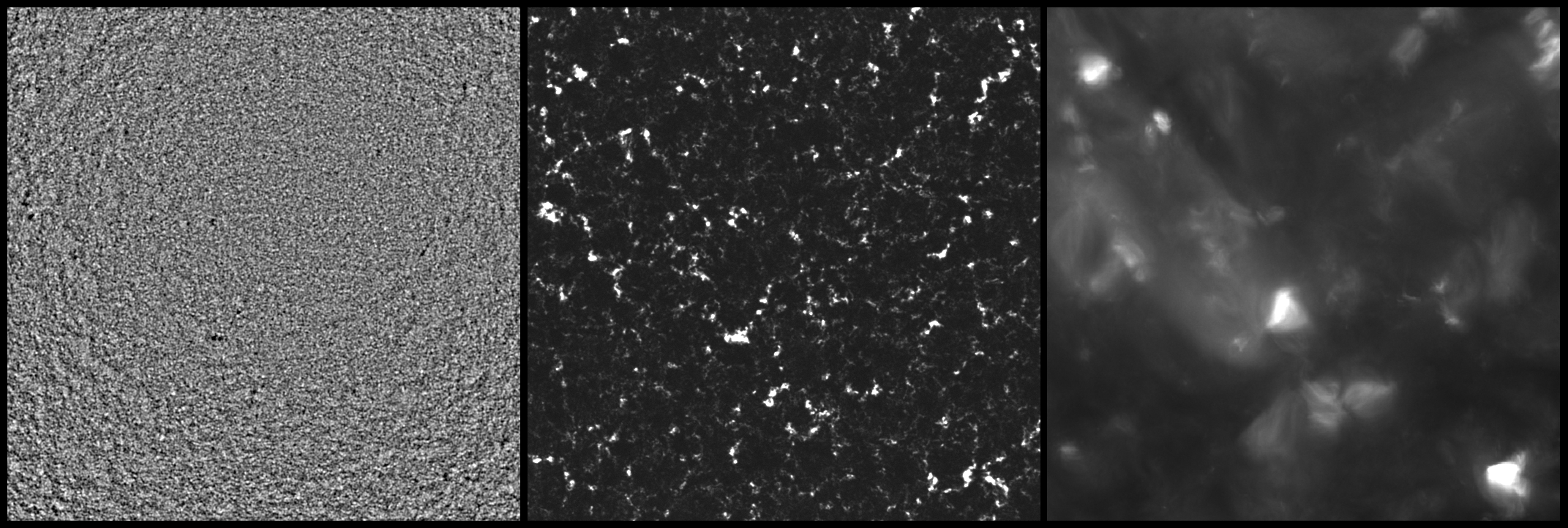}}
    \caption{Region of interest on the 29 November 2018 datacube. The region was localised at the disc centre; the field of view has dimensions of $353\arcsec{} \times 353\arcsec{}$.
    \textbf{Left}: 5-hour mean of the HMI Dopplergrams, where black areas are considered as persistent downflows.
    \textbf{Middle}: 5-hour mean of the line-of-sight magnetic induction  $\normBpar{}$.
    \textbf{Right}: 5-hour mean of AIA 193~\angst{} filtergrams.}
    \label{dop_B}
\end{figure*}

 Recently, the Lagrangian coherent structures \citep[LCS; see, e.g.][]{Chian2019, Chian2020} detection method was applied to locate repulsion and attraction regions as well as the shear and swirling of particle motions on the solar surface. With such a contribution, they highlight that supergranular cells are interconnected by ridges of repelling LCSs that facilitate the formation of vortices and magnetic concentration in the valley of the repelling LCS. The attractive LCS reveals the location of sinks of photospheric flows at supergranular junctions associated with persistent vortices and intense magnetic flux.

The downflows are important ingredients of the convective motions. Until now, they had only been studied  only in small fields of view and with a short cadence. The only exception known to us is the study by \citet{Duvall2010}, who studied the vertical component of supergranulation. The downflows at mesogranular downdraft boundaries act as `collapsars'. Small granules vanish and excite the upward-propagating waves \citep{Rast95, RAST99, Skartlien2000a, Skartlien2000b}. 

The role of surface layers in the global convective simulations was recently discussed by many studies. For instance \citet{Nelson2018} show that the introduction of the near-surface small-scale downflows into the global 3D simulation changes the convective driving motions throughout the convection zone. In particular, the coalescence of downflow plumes into giant cells at larger depths linked to the self-organisation of the near-surface plumes provides a new approach towards the convection conundrum.

Today, the main difficulty seems to be the overestimation of the convective velocity in the convection zone by forward models when forward and inverse models are compared. On the other hand, \citet{Hotta2019} show that the surface region has an unexpectedly weak influence on the deep convection zone and does not resolve the problem of the high convective velocity in the deep solar interior of the state-of-the-art forward models. Nevertheless, \citet{Hotta2019} indicate the possibility of an unknown influence of the unresolved small-scale turbulence.

In this paper, we investigate the properties of near-surface downflows through the observations of the Solar Dynamics Observatory \citep[SDO;][]{Pesnell2012} and Hinode \citep{Kosugi2007} satellites and finally a 3D MHD simulation. In Section~\ref{sec:observ_sim}, we describe the data selection and reduction. The analysis of the persistent downflows is presented in Section~\ref{sec:analysis}. Section~\ref{sec:link_downflows_vortex} is devoted to the potential link between vortex and the persistent downflows. The non-detection of a vortex comparable in size to observations at the surface of the Sun in the simulation is described in Section~\ref{sec:link_downflows_B}. The discussion and conclusion are given in Section~\ref{sec:results}.

\begin{figure}
    \centerline{\includegraphics[width=0.5\textwidth,clip=]{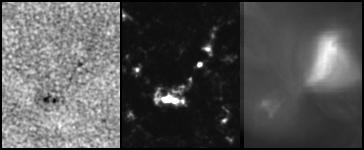}}
    \caption{Details of Fig.~\ref{dop_B} around one of the persistent downflows, with a field of view of $59.5\arcsec \times 73.5\arcsec$. 
    \textbf{Left}: Mean Doppler velocity.
    \textbf{Middle}: Mean $\normBpar{}$.
    \textbf{Right}: Mean AIA 193~\angst{} filtergram.}
    \label{dop_B_AIA}
\end{figure}

\begin{figure}
    \centerline{\includegraphics[width=0.3\textwidth,clip=]{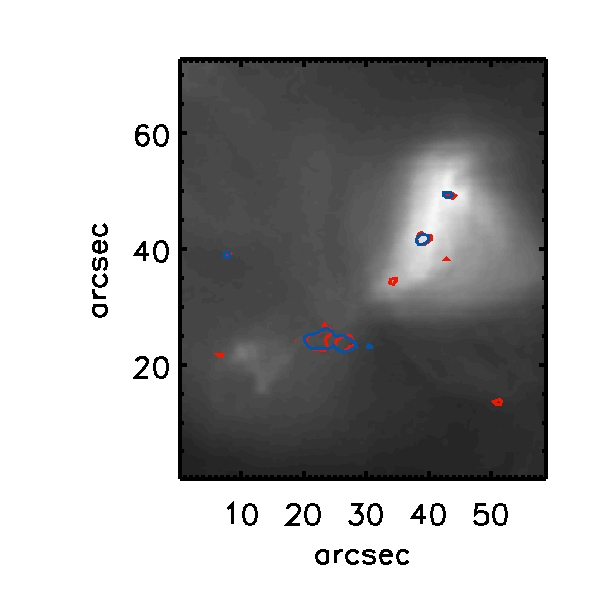}}
    \caption{Coalignment of panels from Fig.~\ref{dop_B_AIA}.
    \textbf{Background}: Mean AIA 193~\angst{} filtergram.
    \textbf{Blue contours}: Mean $\normBpar{}$ level=120 Gauss.
    \textbf{Red contours}: Mean Doppler velocity  level=-0.2~k\mps{} 
    Red and blue contours largely overlap.}
    \label{dop_B_aia_det}
\end{figure}

\section{Observations and simulation}
\label{sec:observ_sim}

To study the properties of the medium-scale surface downflows, we used two 24-h SDO Doppler observations, covering  29 November 2018, and also  2--3 November 2010. Hereafter, we refer to the latter series as the 'Requerey's vortex', where a strong vortical downflow was already studied by \citet{Req2018}. The regions of interest were around the disc centre in the quiet-Sun regions, with a small inclination of the solar rotation axis with respect to the observer (the heliographic latitudes of the disc centre were $B_0=1.1$\degree{} in the first case and $B_0=4.25$\degree{} in the second case). The original pixel size of images were 0.5040\arcsec{} and 0.5042\arcsec{}, respectively, and a time step of 45 seconds. Mainly, we used the observations from the  Helioseismic and Magnetic Imager \citep[HMI;][]{Schou2012}. The 24-h sequences were aligned, corrected from the satellite motions and the limbshift, and finally tracked with the differential rotation. From the full-disc Dopplergrams we extracted a region around the disc centre having $701 \times 701$ pixels on a side $\left( 353\arcsec{} \times 353\arcsec{} \right)$. The tracked Doppler datacube was filtered in the \kw{} domain with a threshold phase velocity of 6~k\mps{}. Line-of-sight HMI magnetograms capturing the longitudinal magnetic induction \Bpar{} were used for both sequences and processed similarly to the Dopplergrams, except for the application of the velocity filter. For the 2018 series, we also used ultraviolet  filtergrams from the Atmospheric Imaging Assembly \citep[AIA;][]{Lemen2012} at a wavelength of 193~\angst{}. The AIA sequence recorded at the original pixel size of 0.6\arcsec{} was mapped to the HMI frame and tracked to coaling with HMI sequences. 
 
We also used datasets from the Solar Optical Telescope \citep[SOT;][]{Tsuneta2008} on-board the Hinode mission spacecraft. 
 The observations were recorded continuously on 4 September 2009, from 7:34:34 to 10:16:34 UT. For our study, we used observations of \ion{Fe}{I} at $\lambda=5250$~\angst{} from the Hinode/SOT-NFI (Narrow band Filter Imager), where the spectral line was scanned at five wavelength positions along the line profile. The Doppler shift at the disc centre gives us the vertical (radial) flow component $v_z$, the Stokes $V$ and hence the vertical component of the magnetic induction $B_z$.  The pixel size of images is 0.16\arcsec{}  and the time step is 60 seconds. After the flat field and dark corrections, the images were coaligned and filtered for $p$-modes in the \kw{} space with a threshold phase velocity of 6~k\mps{}.We note that arcsecond-kilometre conversion, for each dataset considered, takes the distance between the corresponding instrument and the Sun into account. 

In order to learn about the structure of the flows also below the solar surface, we utilised time-distance local helioseismology \citep{Duvall1993}. This method comprises  tools that measure and analyse the travel times of the waves propagating throughout solar interior. 

Travel times of the waves may be measured via cross-correlations of the signals at spatially different points on the solar surface. Doppler shifts of the photospheric absorption lines contain clear indications about solar oscillations. We therefore utilised HMI Dopplergrams and measured travel times using a set of filters, averaging geometries and distances. Since we are dealing with the quiet-Sun region, we measured the travel times using linearised \cite{GB04} method. 

These travel times were then inverted for flows using multi-channel subtractive optimally localised averaging method \citep[MC-SOLA; see][]{Jackiewicz_2012} when involving the cross-talk minimisation \citep{Svanda_2011, Korda_2019}. When discussing the flows, thanks to the cross-talk minimisation our methodology allows one to not only infer the horizontal velocities ($v_x$ in the zonal direction and $v_y$ in the meridional direction), but also the vertical component $v_z$. The inferred velocity vector $\vec{v}$ is a function of the horizontal position $\vec{r}$ and height $z$. Horizontal velocities are strong perturbers that may be successfully retrieved to the depth, whereas the vertical velocity is a weak perturber and only inversions in the first 1~Mm of depth are possible with the signal-to-noise ratio larger than one when averaging the data over 24 hours or so. 

Therefore, except for the surface, the vertical velocity was reconstructed from the horizontal components by integrating the equation of continuity:
\begin{equation}
    v_z \left( \vec{r}_0, z_0 \right) = \frac{-1}{\rho \left(z_0 \right)} \int \limits_0^{z_0} {\rm d}z\, \vec{\nabla}_h \cdot \left[\rho \left(z\right)\, \vec{v}_h \left(\vec{r}_0, z\right) \right],
    \label{eq:vz}
\end{equation}
where $\vec{\nabla}_h \cdot \vec{v}_h$ indicates the horizontal part of the velocity divergence. The integration does not yield the surface $v_z \left(z=0 \right)$, where we directly use the inverted values by helioseismology. The surface inversions for plasma flows were validated against the granule-tracking inferences by \cite{SvandaRoudier_2013}. 

For the deeper layers, we performed a set of inversions targeted such that the vertical sampling was 2~Mm, starting at the depth of 1~Mm and ending at 25~Mm. We basically followed the vertical sampling given by \cite{Greer2016}. With the depth, the extents of the averaging kernels defining the effective resolution increased monotonically, reaching 10~Mm at 1~Mm depth, 34~Mm at 13~Mm depth and finally 58~Mm at 25~Mm depth. The vertical extent defining the vertical resolution increased as well, having 1~Mm at the depth of 1~Mm, 2.9~Mm at the depth of 13~Mm and finally 8.7~Mm at the depth of 25~Mm.  

\begin{figure*} 
 \centerline{
   \includegraphics[width=0.45\textwidth,clip=]{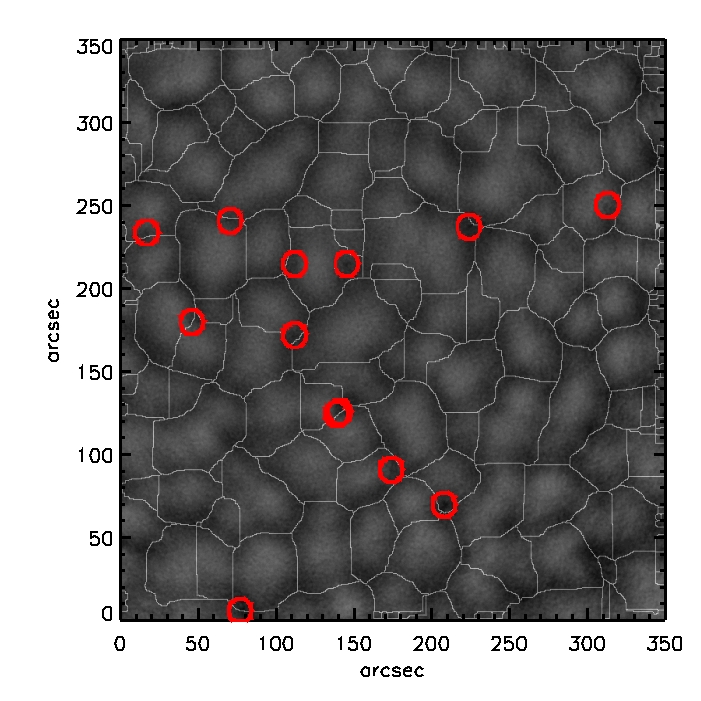}
   \includegraphics[width=0.45\textwidth,clip=]{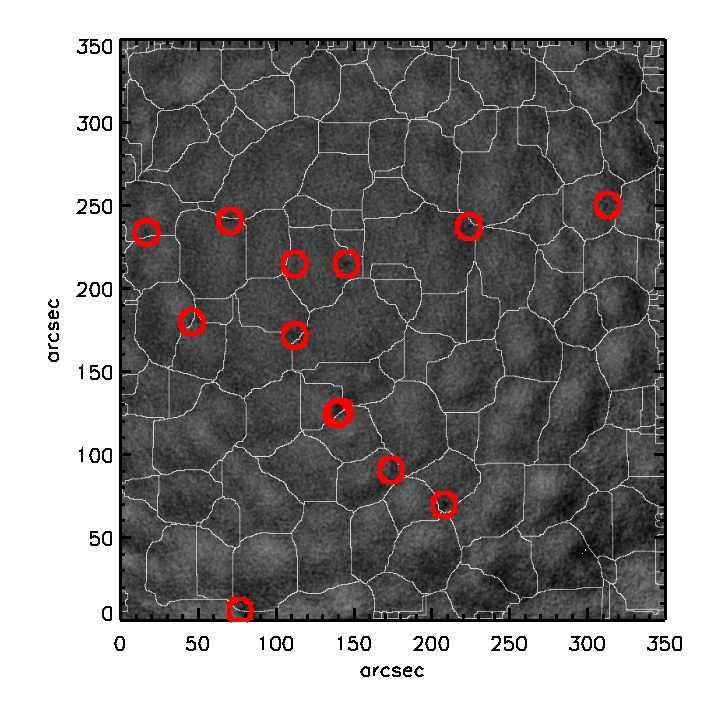}}
   
 \centerline{
   \includegraphics[width=0.45\textwidth,clip=]{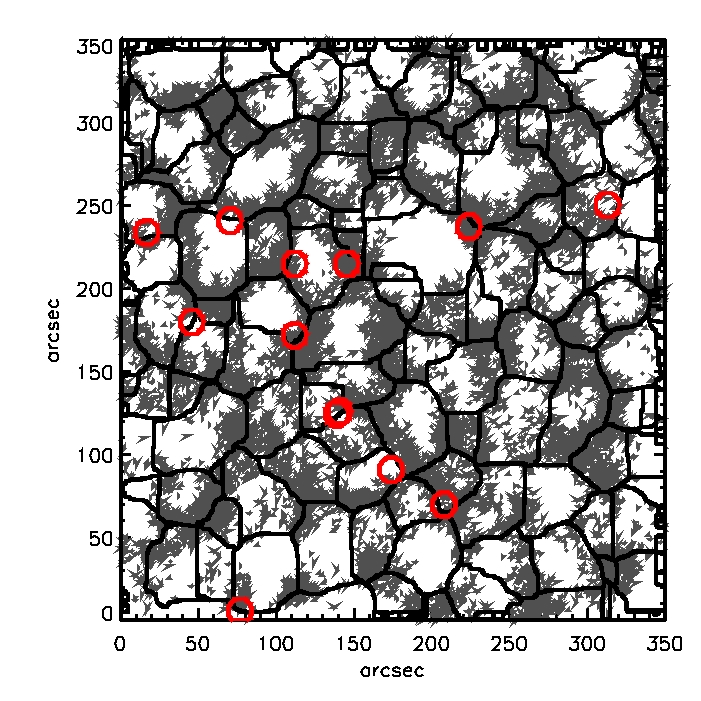} 
   \includegraphics[width=0.45\textwidth,clip=]{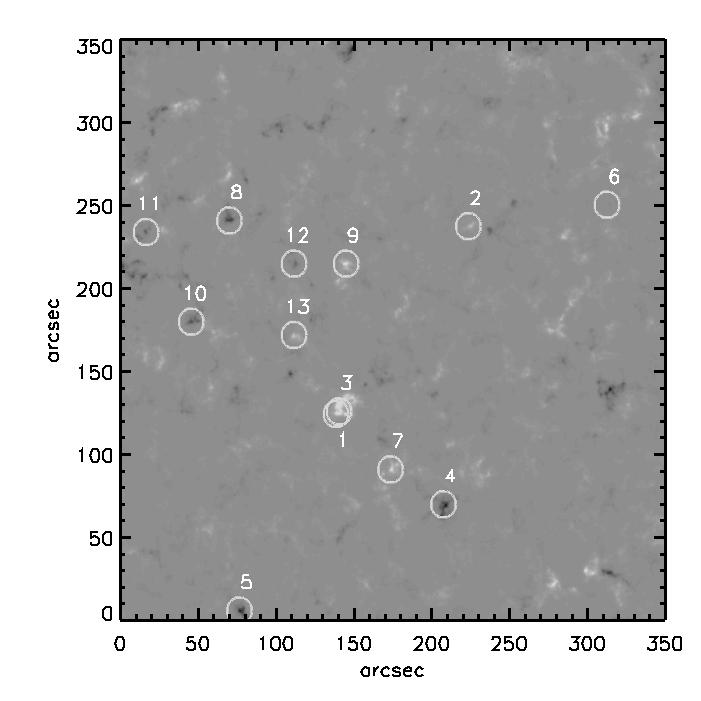}}
     
   \caption{
     {\bf Top left:} 24-hour divergence field measured with the LCT applied to Dopplergrams. The supergranule boundaries detected by the watershed method are shown as white network. The circles represent the location of the 13 studied a persistent downflows. These persistent downflows are located in convergent flows and more particularly where several supergranules meet.
     {\bf Top right:} 24-hour mean Doppler map overlapped with the 13 studied persistent downflows (circles) and supergranule boundaries. Downflows are visible as dark structures in each circle. Far from the disc centre, in the lower right part of the figure, some large darker regions are visible. These are caused by the horizontal velocities which become a dominant contribution due to the projection effects.
     {\bf Bottom left:} Corks position (grey points) after 24-hour diffusion by horizontal flows. Their locations are, in accordance with supergranule boundaries, found by the watershed method (superimposed white network).
     {\bf Bottom right:} 24-hour mean magnetic field and the location of the 13 studied persistent downflows (circles).
    }
 \label{diverg}
 \end{figure*}

\begin{figure*}
    \sidecaption
    \centerline{\includegraphics[width=0.55\textwidth,clip=]{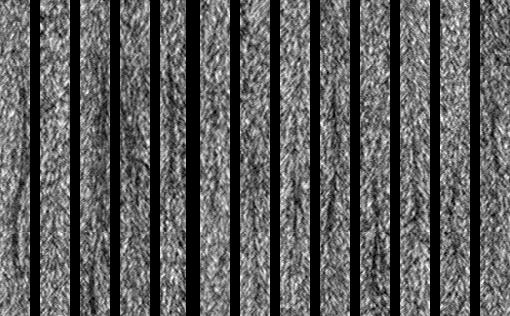}}
    \centerline{\includegraphics[width=0.55\textwidth,clip=]{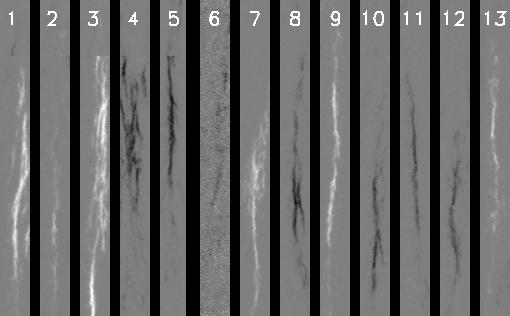}}
    \caption{Doppler (top) and magnetic (bottom) temporal cuts centred on the 13 selected persistent downflows.
    The time (vertical axis) stands for 24 hours, the spatial axis (y cut, horizontal) is 15\arcsec{} for each column and the numbers indicate the labels of the selected downflows corresponding to the bottom right of Figure~\ref{diverg}. The persistent downflows move across the surface but are visible in the cuts at least from 3.5~h to 20~h. 
    }
    \label{Cuts_dop1}
\end{figure*}

\begin{figure}
    \centerline{\includegraphics[width=0.5\textwidth,clip=]{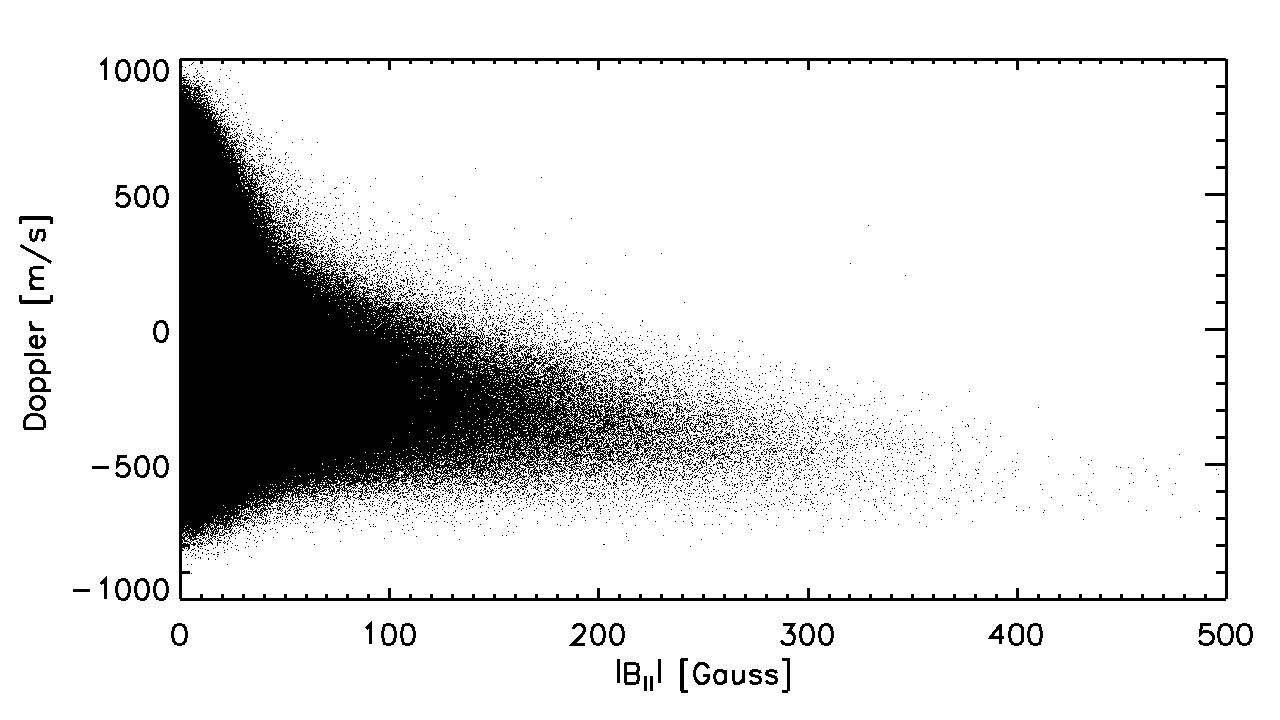}}
    \caption{The scatter plot of the dependence of Doppler velocity and $\normBpar{}$ obtained for the SDO data on 29 November 2018.}
    \label{dop_B_rel}
\end{figure}

\section{Analysis of persistent downflows with HMI, Hinode, and simulation data}
\label{sec:analysis}

The first detection of persistent (long-lived) downflows surviving for at least 4 hours is described in \citet{Duvall2010} (their Figs.~1a, 1c). The results were obtained by the analysis of the Dopplergrams and magnetograms from SOHO/MDI \citep{Scherrer1995} with a resolution of 1.2\arcsec{}. \citet{Duvall2010} selected a 4-hour temporal window to average Dopplergrams at the disc centre to follow the evolution of supergranular cells. They tested a possible contamination of the Doppler signal by magnetic fields but concluded that there was a rather small contribution from the magnetic field. Their simultaneous 4-hour average of Dopplergrams and magnetograms showed a correlation between the persistent downflows and the magnetic induction. We note that the persistent downflows were called apparent downflows in their paper.

For our study, we selected a 24-hour sequence of disc centre quiet-Sun Dopplergrams and magnetograms observed on 29 November 2018 with the SDO/HMI. The spatial resolution of these observations was 1.0\arcsec{}.
To mimic the downflow detection like  \citet{Duvall2010}, we used the 5-hour average Dopplergrams. Figure~\ref{dop_B} (left) convincingly shows persistent downflows as localised black areas. The large amplitude of the negative vertical velocity indicates that the downflow survived 24-h averaging. These downflows with sizes  around 2--3\arcsec{} and velocity amplitudes between $-0.25$ and $-0.72$~k\mps{} spatially correlate with the magnetic field in location (see the middle panel of Fig.~\ref{dop_B}) and also with the regions of larger activity in the corona (images from AIA 193~\angst{}). Figure~\ref{dop_B_AIA} shows a zoom-out of the larger downflows from Figure~\ref{dop_B}. Figure~\ref{dop_B_aia_det} displays the persistent downflows (red) and $\normBpar{}$ (blue) at the exact locations relative to the coronal loop. In this example the persistent downflows appear to correlate with the coronal-loop anchor in the photosphere, indicating a link between the persistent downflows and the coronal activity.

The inspection of the 3D $\left(x, y, t\right)$ Dopplergram datacube from 29 November 2018 allowed us to detect 13 persistent downflows with a minimum duration of 3 hours (see below). This gives a rate of occurrence of $2\times 10^{-4}$ cases per Mm$^{2}$ and 24~h. This rate is lower than those found by \citet{Req2017} during sequences of 32.0 and 22.7~minutes of $6.7\times 10^{-2}$ per Mm$^{2}$.  These two rates of occurrence are not directly comparable because the first one deals with very long-lasting downflows (a few hours) although the second rate is related to shorter-lasting downflows (30 minutes).

To detect the supergranule boundaries, horizontal velocities were computed with the local correlation tracking technique \citep[LCT;][]{Nov86} applied to the sequence of Dopplergrams. From the derived horizontal velocities, we computed the 24-h average divergence field and the supergranule boundaries using the watershed method \citep{Rou2020}. These boundaries plotted over the divergence field are shown as a white network in Figure~\ref{diverg} (top left). In Fig.~\ref{diverg} (top right) the 24-hour mean Doppler map is overlapped by locations of the 13 studied persistent downflows (circles) and supergranular boundaries. Downflows are visible as dark structures in each circle. The persistent downflows are identified in the convergent flows and more particularly at positions, where several supergranules meet. This is in agreement with the conclusion that vortex flows are usually located at supergranular junction vertices found by \citet{Attie2009, Attie2016}. Using the 24-hour horizontal-velocity fields, we computed the corks diffusion during that period of time. Fig.~\ref{diverg} (bottom left) confirms that the locations of the persistent downflows (circles) at the supergranule boundaries are situated where the corks are accumulate. 

Temporal cuts of the Doppler velocities centred on the 13 persistent downflows are shown in Fig.~\ref{Cuts_dop1} (top). The downflows are visible as dark structures in the cuts and they last for a long time from 3.5~h to 20~h in our examples. However, the downflows move across the surface during their lifetimes which makes it difficult to catch them on single positional cuts. Nevertheless, the Doppler temporal cuts clearly exhibit  their long duration. Some of them are not fully continuous but are visible at the same position. The sizes of these downflows are around 2--3\arcsec{}.
 
Fig.~\ref{Cuts_dop1} (bottom) shows magnetic field temporal cuts centred on the 13 persistent downflows. The correlation between the persistent downflows and the magnetic field is well observed and conforms to the one found by \citet{Duvall2010}. The magnetic field is always observed at the location of the persistent downflows. From our set, only one persistent downflow (number 6) seems to be without a strong magnetic counterpart (see bottom right panel of Fig.~\ref{diverg}).

To quantify the velocity-magnetic field correlation, we plotted the mean Doppler velocity versus $\normBpar{}$ from the 24-hour sequence in Fig.~\ref{dop_B_rel}. In the regions where $\normBpar{} > 60$~G, a linear relationship between the magnetic field and the observed Doppler velocity is most clear as described by \citet{Duvall2010}. The strongest magnetic concentrations are found in the strongest downflows. As we observe with high-resolution HMI data (0.5\arcsec{} per pixel), the positive correlation probably indicates a suppression of the granular flows due to the magnetic field. Our observations thus confirm the hypothesis by \citet{Duvall2010}.

\subsection{Depth structure of the persistent downflows}

The persistent downflows identified in the HMI frame were also localised in the helioseismic datacubes. Unlike full-resolution Dopplergrams or velocity fields obtained by LCT, the helioseismic inferences have a much coarser spatial resolution, 10~Mm effective resolution at the surface at best. Therefore these spatially confined persistent downflows cannot be directly identified in the helioseismic frames. We verified that no clear signal of these confined downflows is present in the travel-time maps. This is due to the fact that the typical wavelength of the $p$-modes and the surface gravity $f$-modes is a few megametres, which is much larger than the horizontal extent of the identified persistent downflows. 

\begin{figure}
    \centerline{\includegraphics[width=0.5\textwidth,clip=]{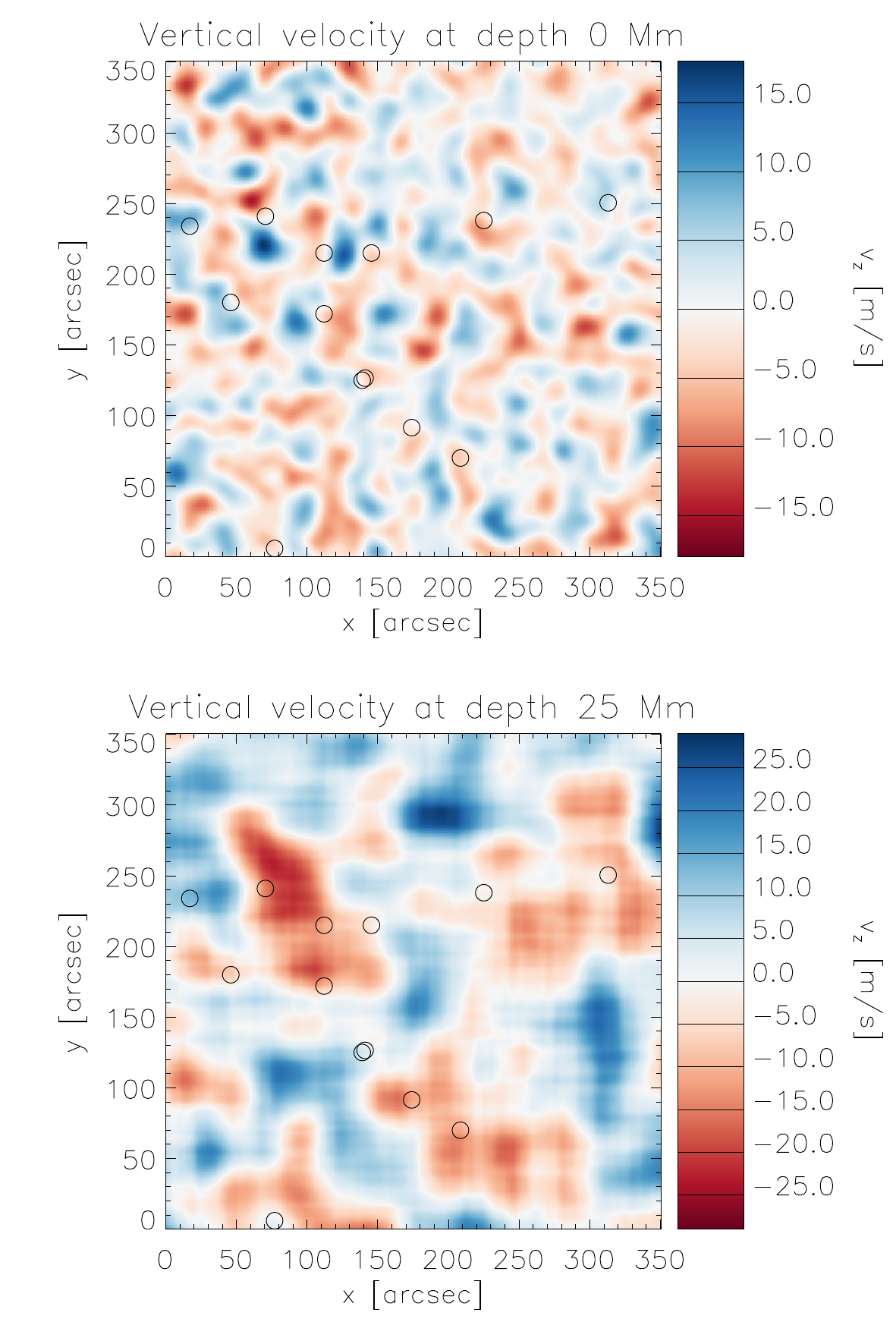}}
    \caption{Locations of the 13 selected downflows (open circles) on vertical velocity map at the surface (top) and at the depth of 25~Mm (bottom). The flows were averaged over 24 hours on the day of 29 November 2018. }
    \label{Vz_loc}
\end{figure}

Hence, the downflows were  localised by their coordinates. In Fig.~\ref{Vz_loc} we see their locations on the vertical-flow maps derived at the surface (a helioseismic inversion) and at the depth of 25~Mm (estimate obtained from integrating Eq.~(\ref{eq:vz})). At the surface frame the locations of the persistent downflows do not seem to be special except that they are mostly located at the edges of supergranules. On the other hand, at a depth of 25~Mm most of them seem to be present at locations of the negative vertical velocity or very close to it. 

\begin{figure}
    \centerline{\includegraphics[width=0.5\textwidth,clip=]{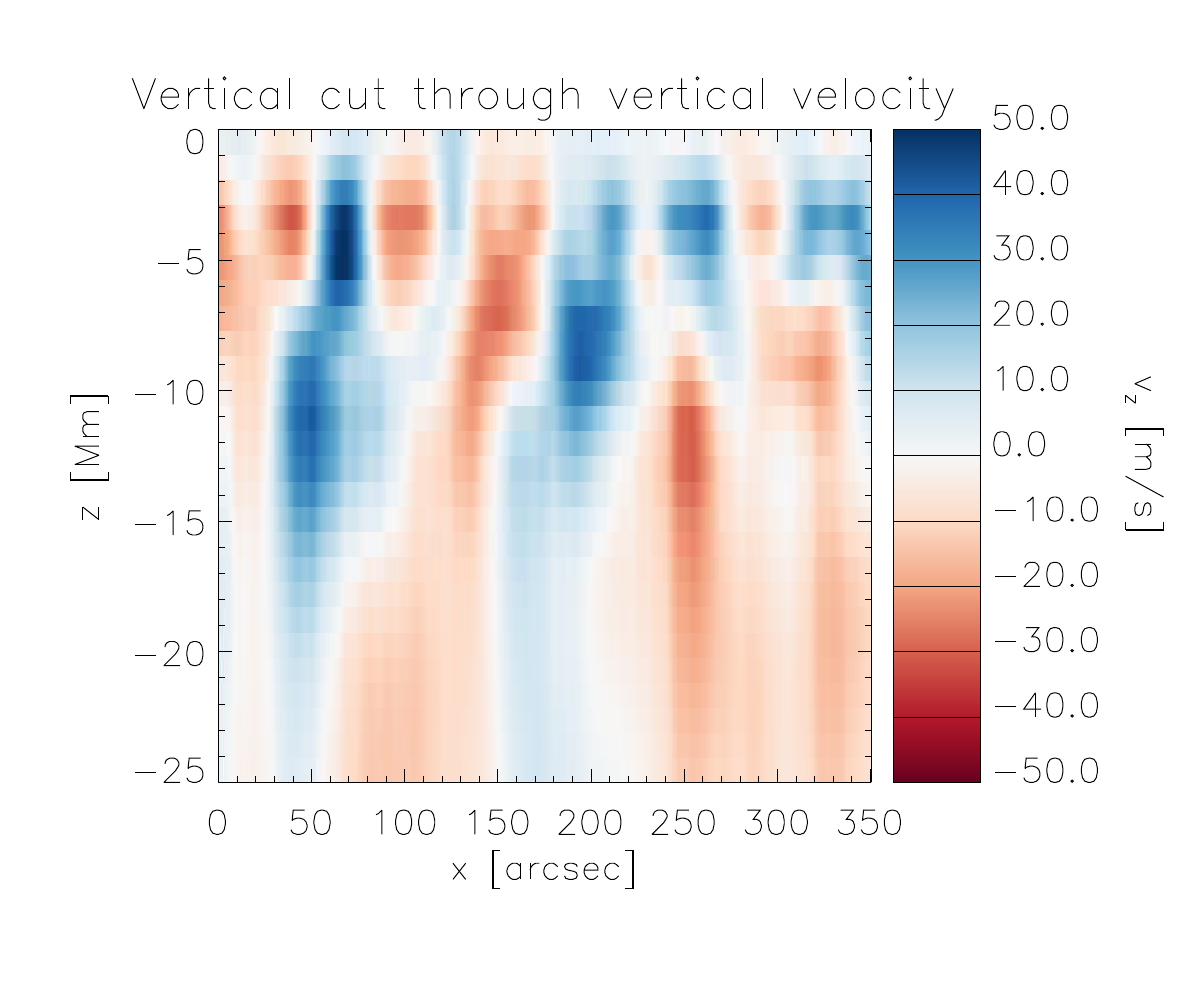}}
    \caption{Example vertical cut through the 3D vertical velocity at \mbox{$y = 210\arcsec{}$}.}
    \label{Vz_cuts_examp}
\end{figure}

Vertical cuts through vertical velocity generally show a complicated structure (see an example in Fig.~\ref{Vz_cuts_examp}). In most cases, it is not possible to track the downflows or upflows through the datacube in  a depth strictly vertically. In some cases the vertical velocities create compact regions that are shifted laterally in between the consecutive depths. Merging of the downflows and separation of the upflows are visible in the cuts. 

\begin{figure*}
    \sidecaption
    \includegraphics[width=0.7\textwidth,clip=]{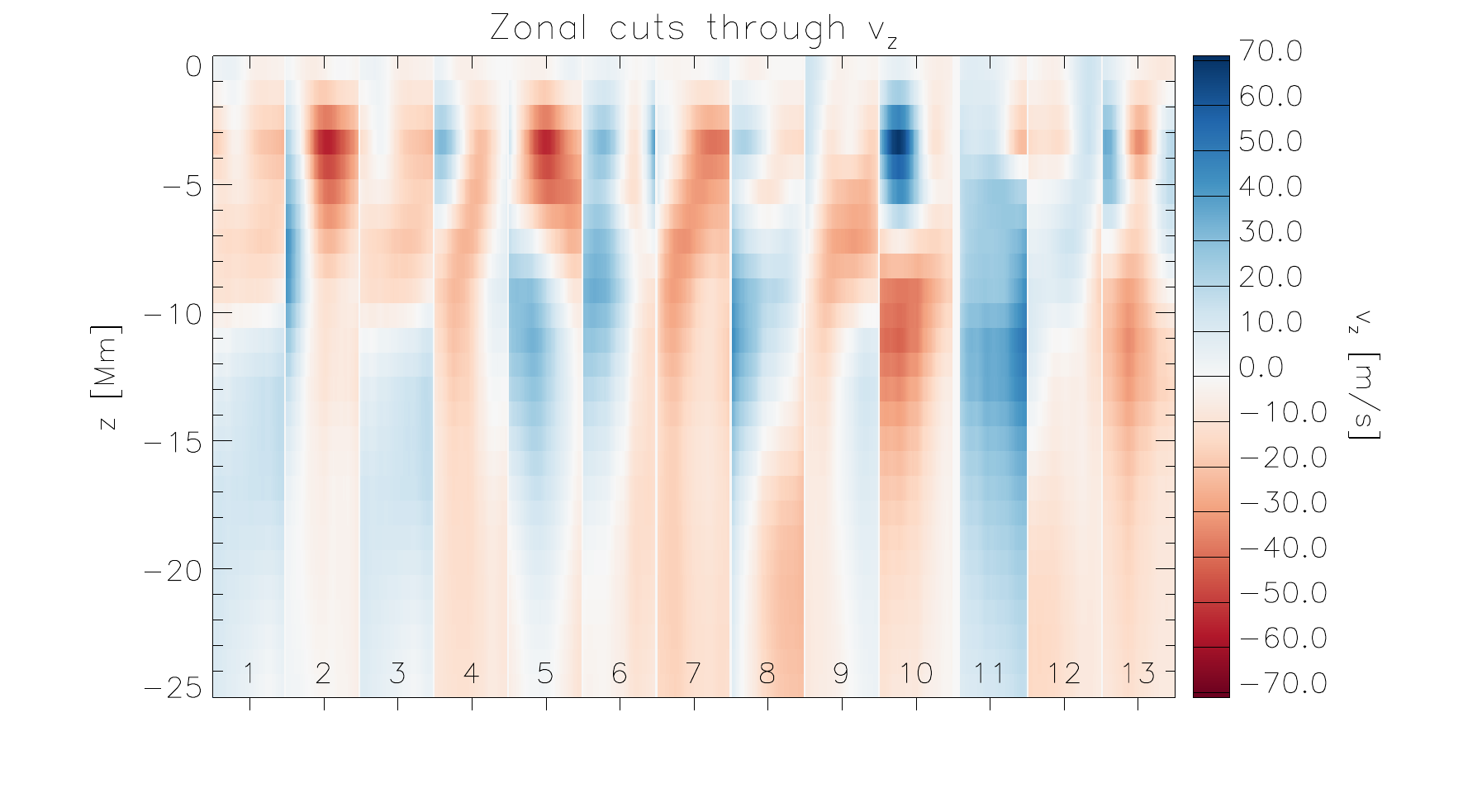}\\
    \includegraphics[width=0.7\textwidth,clip=]{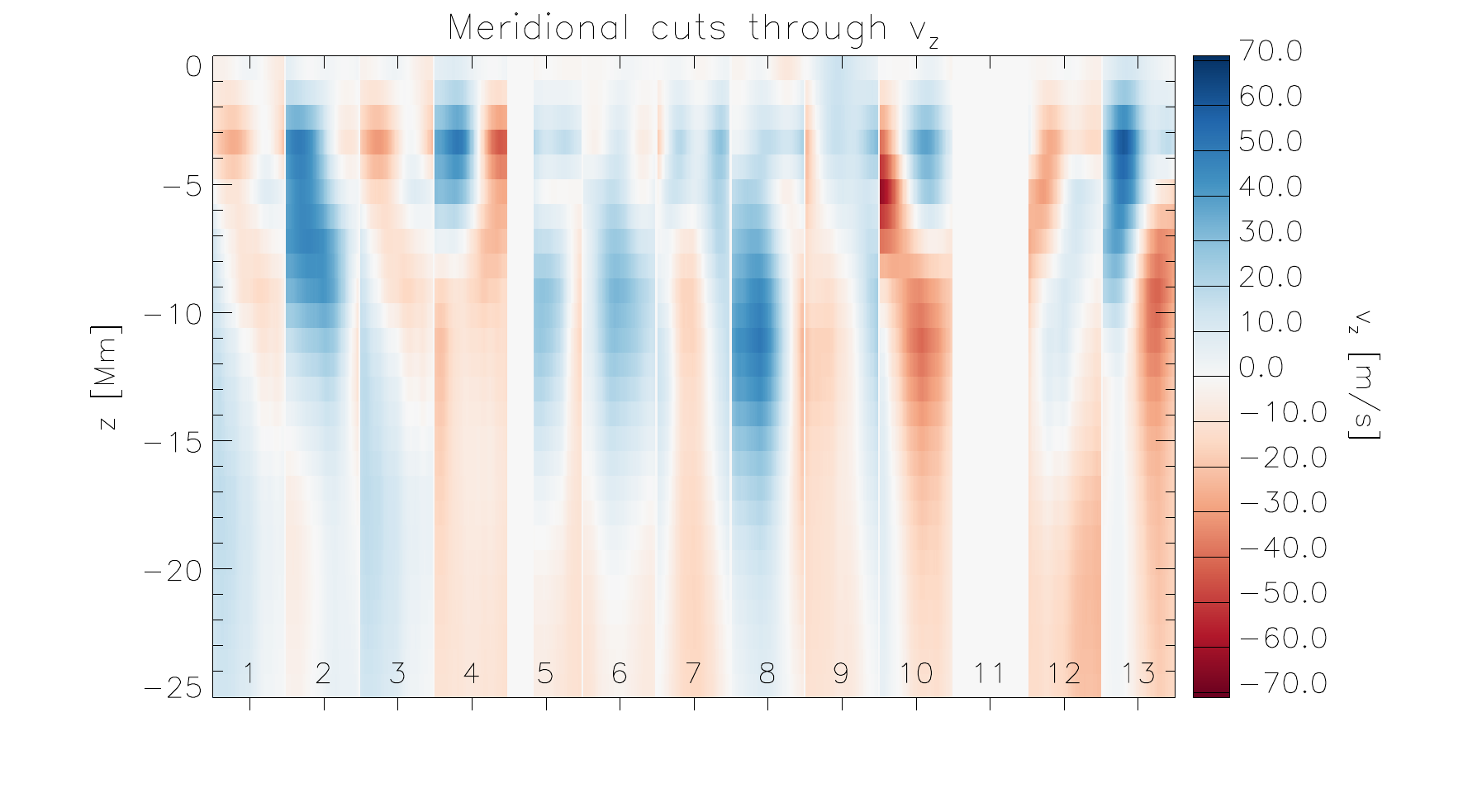}
    %
    \caption{Vertical cuts through the vertical velocity around the 13 selected downflows from 29 November 2018. Around each downflow, a vicinity of 20\arcsec{} was segmented and the cut was plotted in both zonal and meridional directions. }
    \label{Vz_cuts}
\end{figure*}

In this study, we did not focus on a general analysis of downflows, we rather studied the 13 representatives that prevailed for a long time and were associated with the magnetic elements. The vertical cuts through these downflows are seen in Fig.~\ref{Vz_cuts}. As one can see, in most cases the downflow seems to be present deep, in eight out of 13 cases all the way to the bottom of the domain of interest. 

\begin{figure}
    \centerline{\includegraphics[width=0.5\textwidth,clip=]{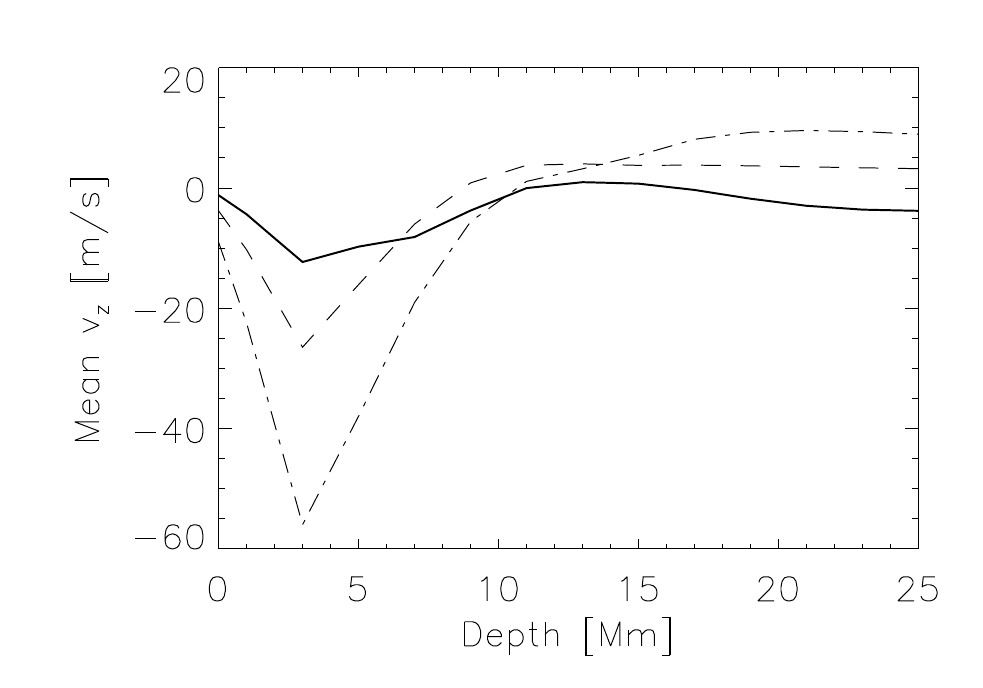}}
    \caption{Average depth profiles of the vertical flows in downflows. The dashed line represents the average over all 183 downflows located in the field of view. The dot-dashed line is an average over the 13 strongest (at the surface) downflows in the region. The solid line then shows the plot for the 13 persistent downflows. }
    \label{mean_vz_cuts}
\end{figure}

These 13 selected persistent downflows are not the only downflows present in the field of view. By doing an automatic search we identified 183 downflows in the region of interest as the local minima of the vertical velocity which, at the same time, exhibited the negative value. From those downflows we selected the 13 strongest to represent a comparative set to the 13 persistent downflows. For each of the three 
sets, we computed the average vertical flow profile in the centre of the downflow. These plots are given in Fig.~\ref{mean_vz_cuts}. We note that our results are only weakly influenced by the downflow changing the position in the coordinate frame, as these positional changes are rather small (about 5\arcsec{}). These positional changes are significantly smaller than the effective resolution of the helioseismic flow maps (about 13\arcsec{} at the surface, the number increases with depth). 

There is an obvious difference between the common and the persistent downflows. The persistent downflows, on average, may have a smaller amplitude both at the surface and in the near-surface layers, but on average the flows remain negative all the way to the bottom of our inversion domain. Compared to that the average over all common downflows tends to be zero at the depth of about 8~Mm and remains negligible further down. The comparative set of the strongest downflows has a much larger magnitude peaking at the depth of about 3~Mm on average, but it reaches zeros at the depth of about 13~Mm  on average and deeper down it even turns positive. 

\subsection{Persistent downflows in 2h30min  Hinode observations}

\begin{figure}
    \centerline{\includegraphics[width=0.3\textwidth,clip=]{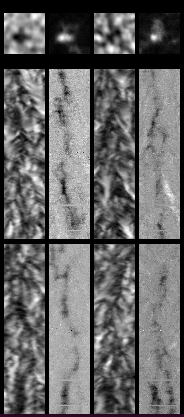}}
    \caption{Two persisting downflows during 2h30min of  Hinode observations.
    \textbf{Top}: Averaged Dopplergrams and $\normBpar{}$ around two different persistent downflows in a field of view of $4.4\arcsec{} \times 4.4 \arcsec{}$.
    \textbf{Middle}: Cuts around the persistent downflows in the $\left(y ,t \right)$ plane of the Dopplergram and $\normBpar{}$ datacubes. The dimensions of the arrays are ($4.4\arcsec{} \times$ 2h30min).
    \textbf{Bottom}: The same in the $\left(x ,t \right)$ plane. }
    \label{cuts_hin}
\end{figure}

\begin{figure}
    \centerline{\includegraphics[width=0.5\textwidth,clip=]{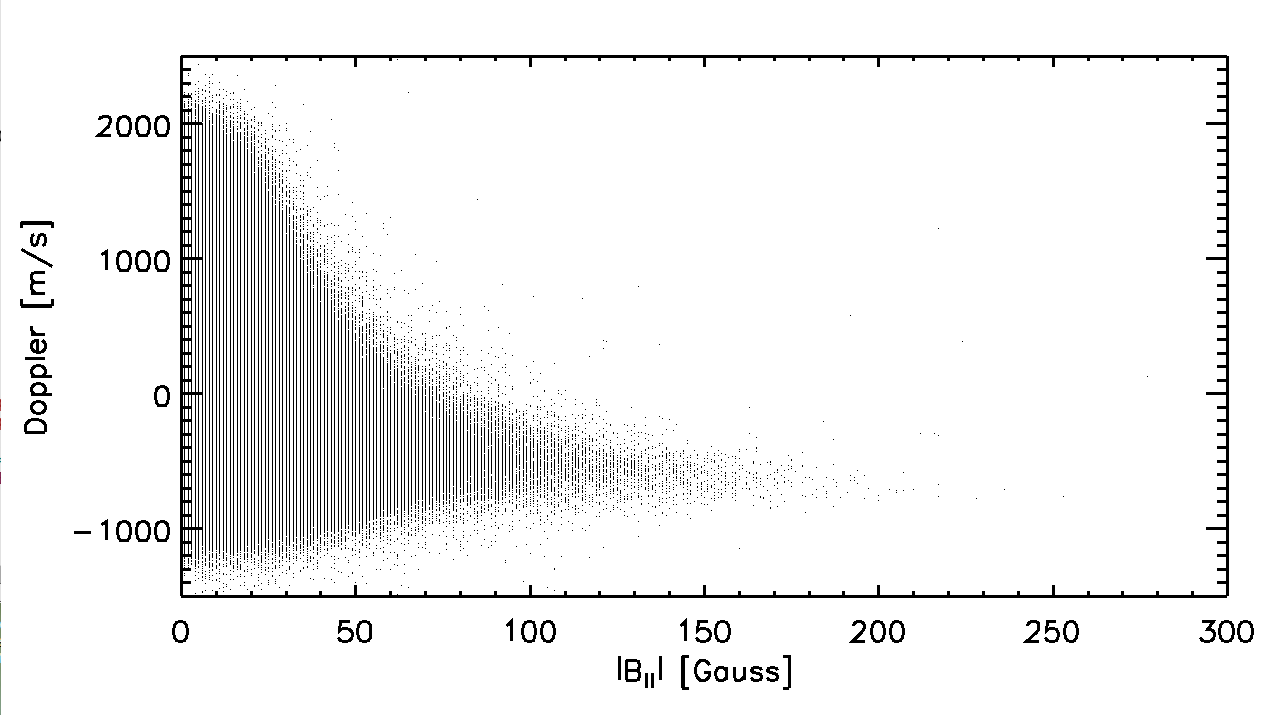}}
    \caption{Correlation between $\normBpar{}$ and the Doppler velocity of the mean data 2h30min of the mean Hinode data.}
    \label{mean_B_dop}
\end{figure}

As a complementary series, we analysed the 2h~30min series of Dopplergrams (based on \ion{Fe}{I} at $\lambda=5250$~\angst{} line) and the longitudinal magnetic field. SOT/NFI on-board Hinode has a spatial resolution of 0.35\arcsec{}. Two examples of persistent downflows found in the datacube sequences are displayed in Figure~\ref{cuts_hin}. There, in the top row, we see an average of the Doppler and $\normBpar{}$ over 2h30min.  In the bottom row, the cuts in the $\left(x,t \right)$ and $\left(y,t \right)$ planes through Dopplergram and $\normBpar{}$ datacubes are plotted at the centres of the two persistent downflows. Here, the persistent downflows are clearly associated with the presence of the magnetic field. The relationship between the magnetic field and the downflows (Fig.~\ref{mean_B_dop}) is identical to that obtained with the SDO data. It is important to note that the amplitude of the magnetic field is lower in the Hinode data, which is due to the smaller field of view and a quieter region.

\section{Link between persistent downflows and potential vortex}
\label{sec:link_downflows_vortex}

\begin{figure}
    \centerline{\includegraphics[width=0.5\textwidth,clip=]{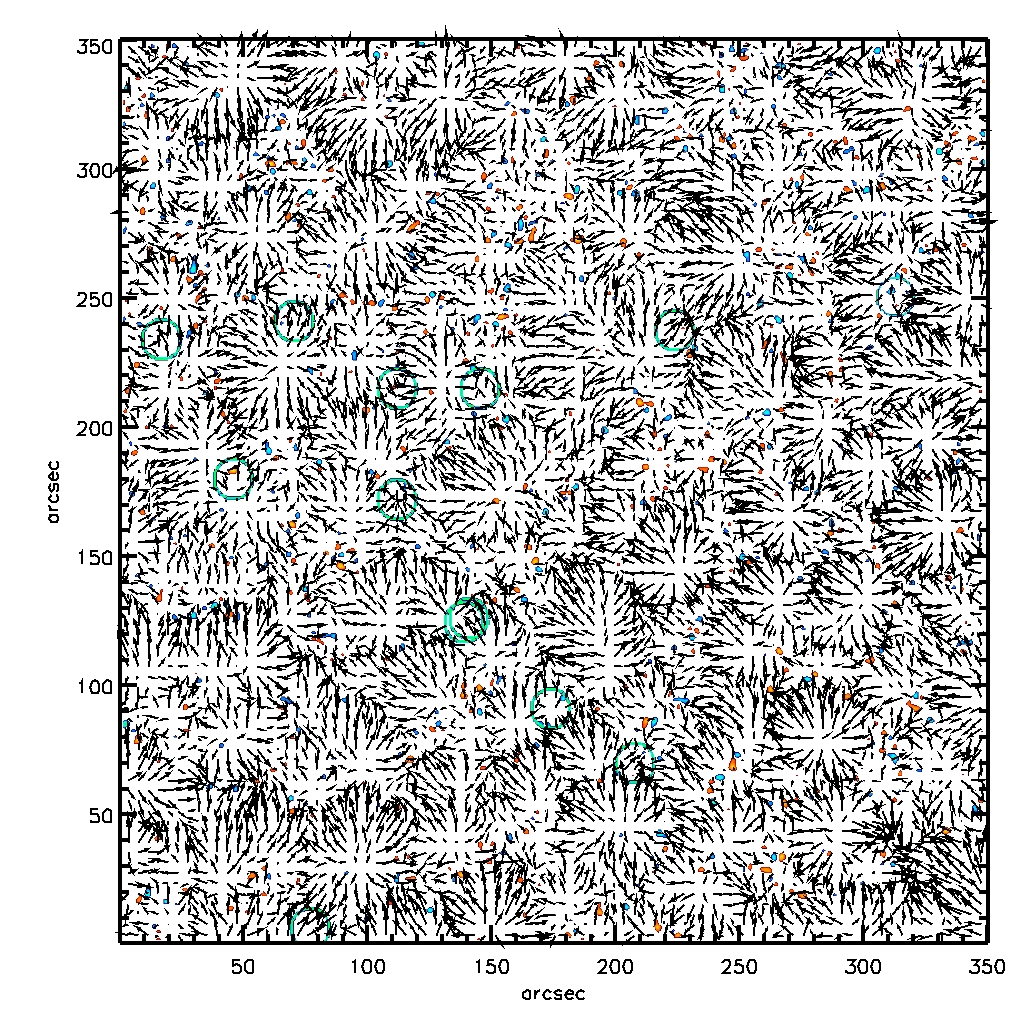}}
    \caption{$\Gamma_1$ computed from 24-h mean velocity overlapped by horizontal velocity field. Blue  and red represent counter clockwise and  clockwise velocities respectively. The locations of the 13 studied persistent downflows are indicated by circles. }
    \label{Gam_29nov}
\end{figure}

\begin{figure*} 
 \centerline{
   \includegraphics[width=0.35\textwidth,clip=]{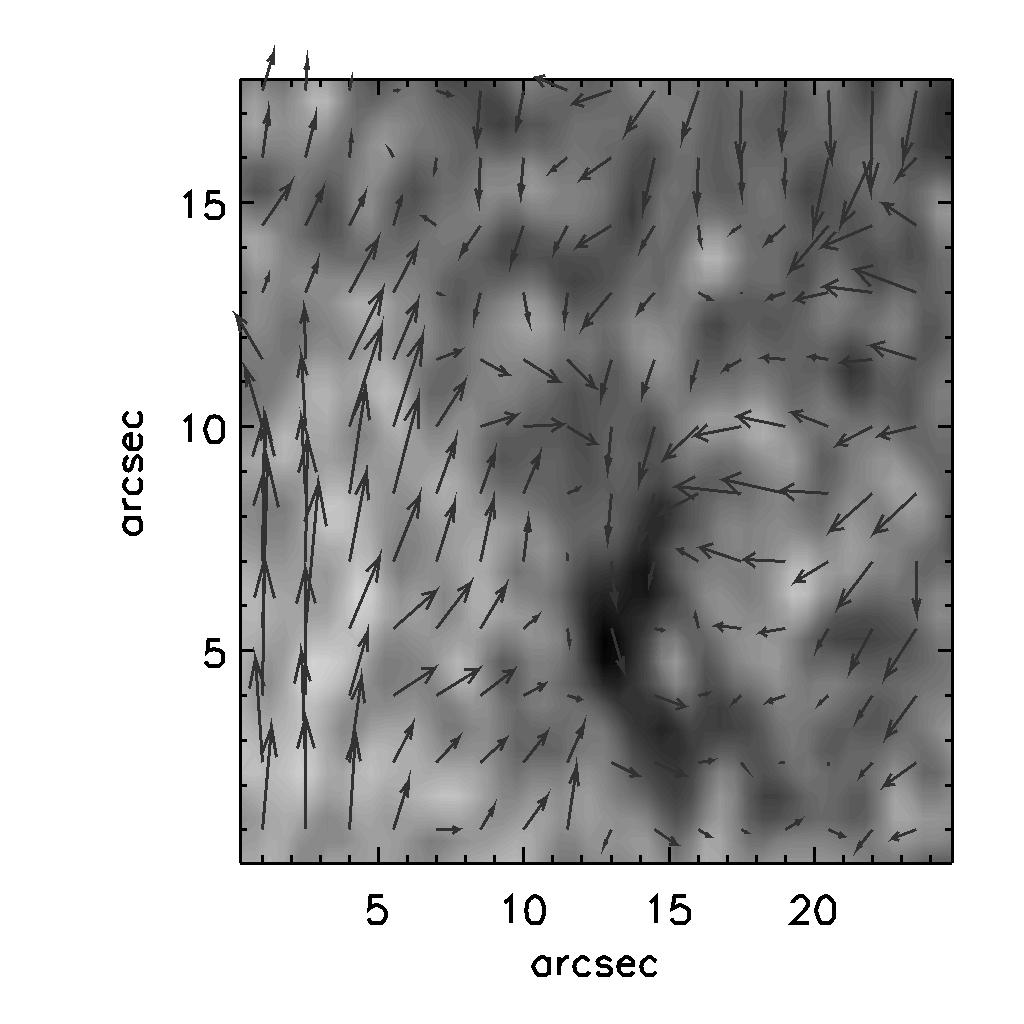}
   \includegraphics[width=0.35\textwidth,clip=]{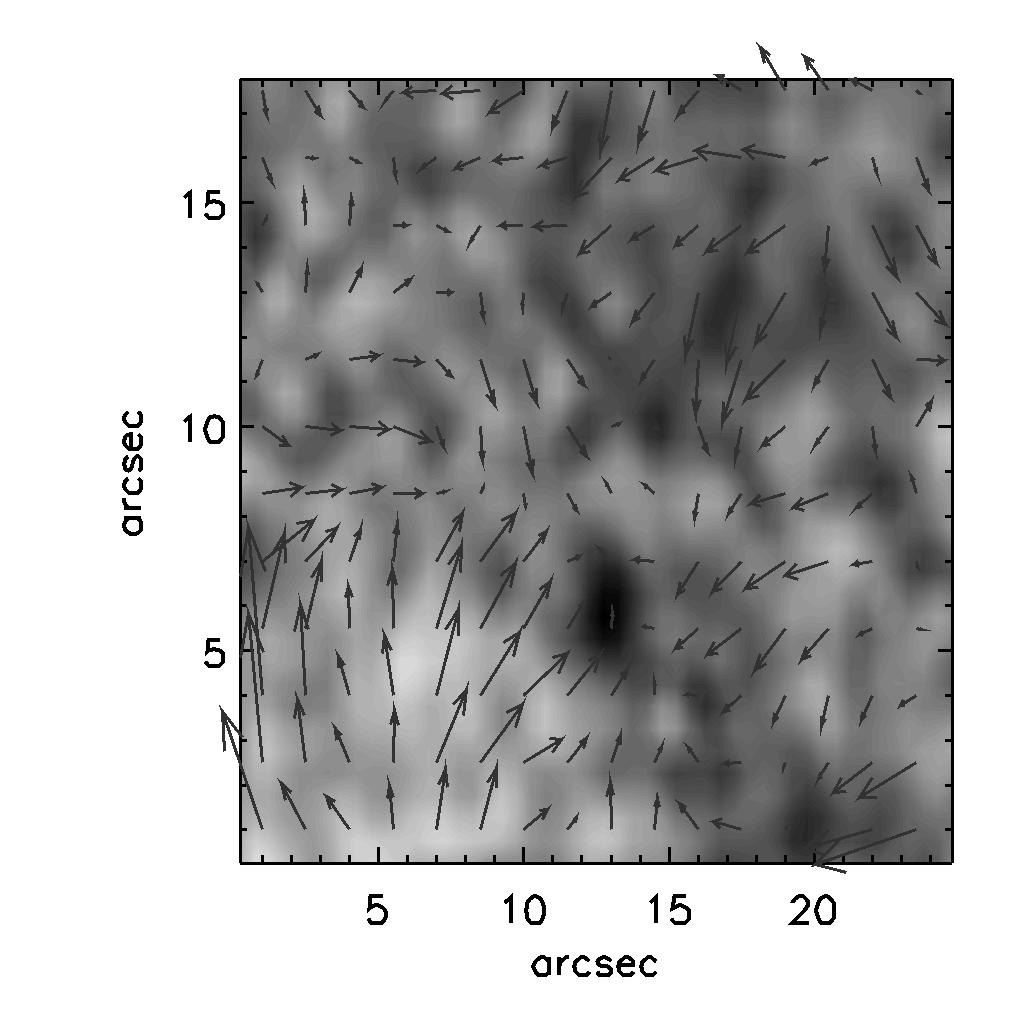}}
   
 \centerline{
  \includegraphics[width=0.35\textwidth,clip=]{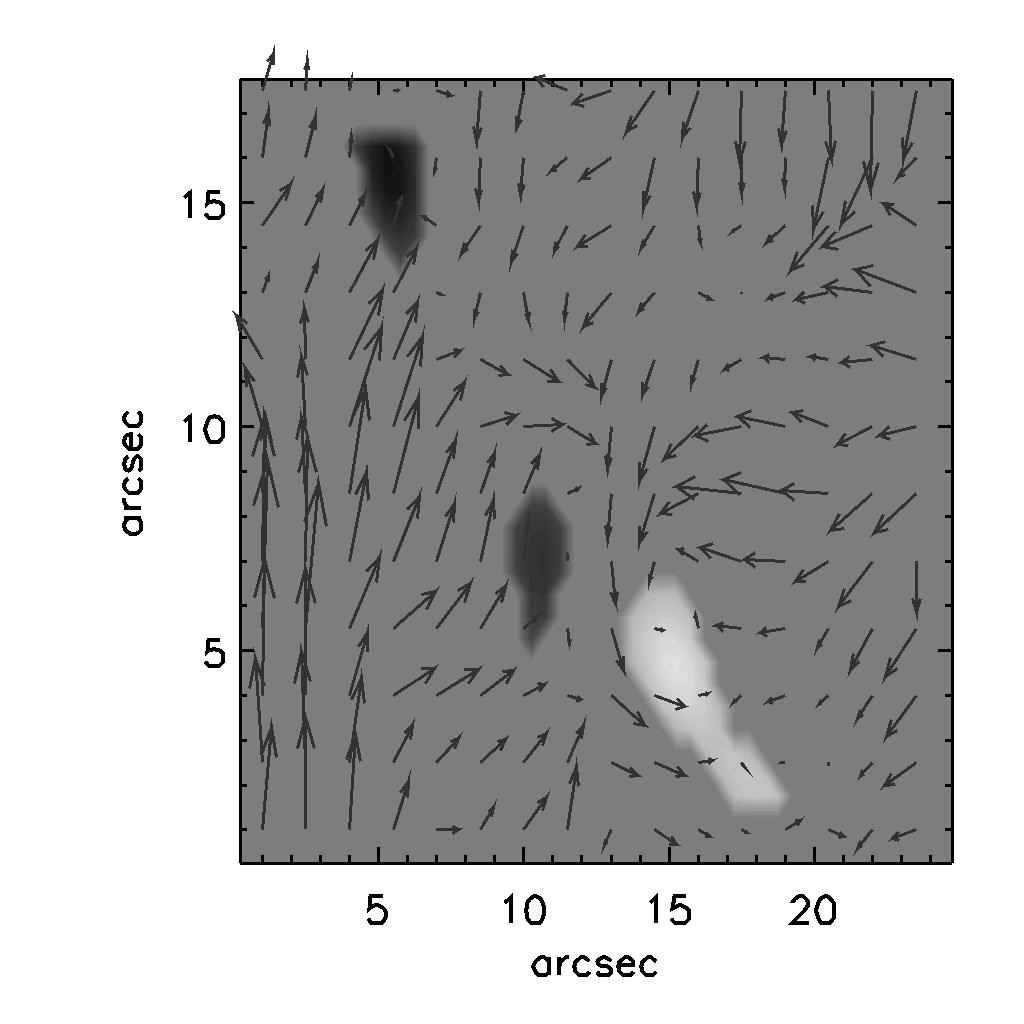}
  \includegraphics[width=0.35\textwidth,clip=]{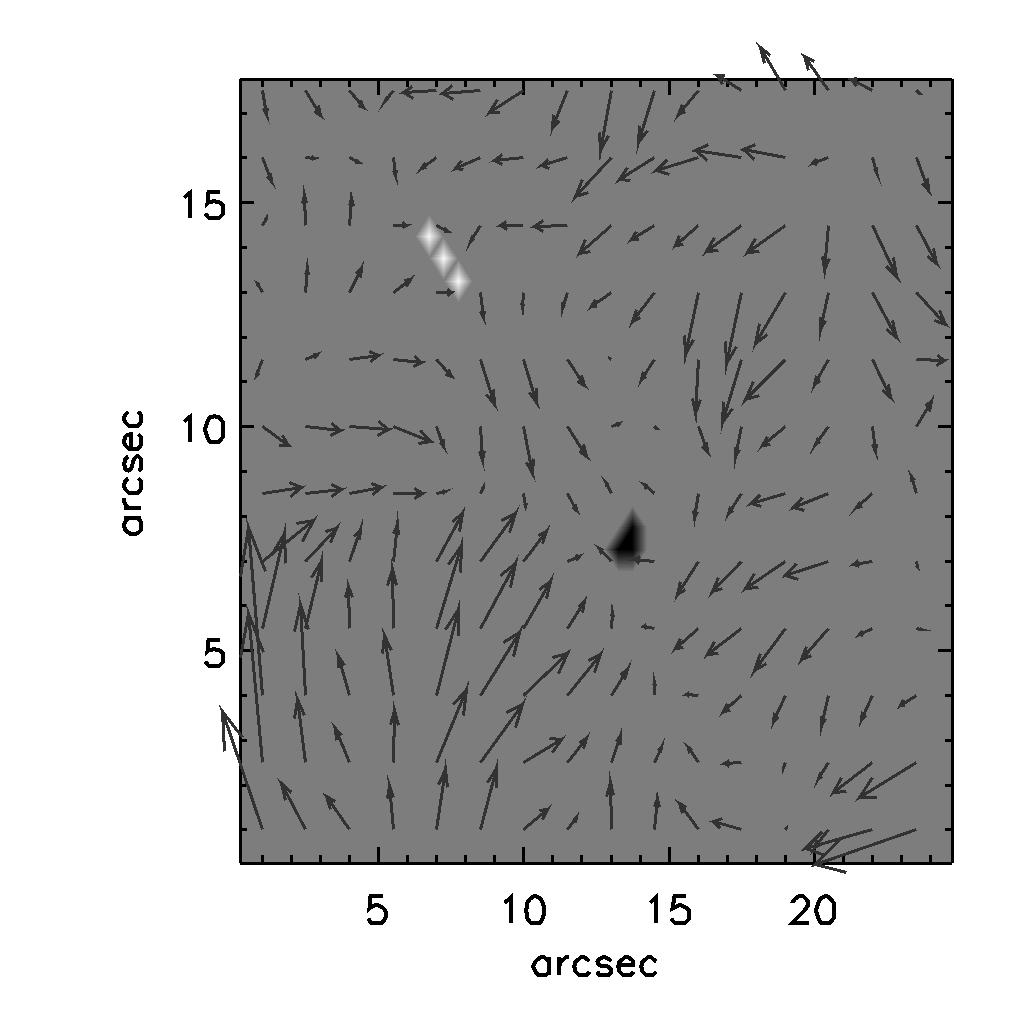}}
     
 \caption{
   {\bf Top:} Mean Doppler downflow number~5 from time 11.4~h to time 15.2~h (left) and from time 15.2~h to 19.7~h (right) relative to the beginning of the sequence, overlapped with the mean horizontal velocities.
   {\bf Bottom:} Vorticity characterised by the $\normG >0.4$ parameter (downflow number~5) from time 11.4~h to 15.2~h (left) and from time 15.2~h to 19.7~h (right), overlapped with the mean horizontal velocities. White and black represent counter-clockwise $\Gamma_1$ and  clockwise velocities, respectively.
 }
 \label{down5_gam}
\end{figure*}

\begin{figure*} 
 \centerline{
   \includegraphics[width=0.35\textwidth,clip=]{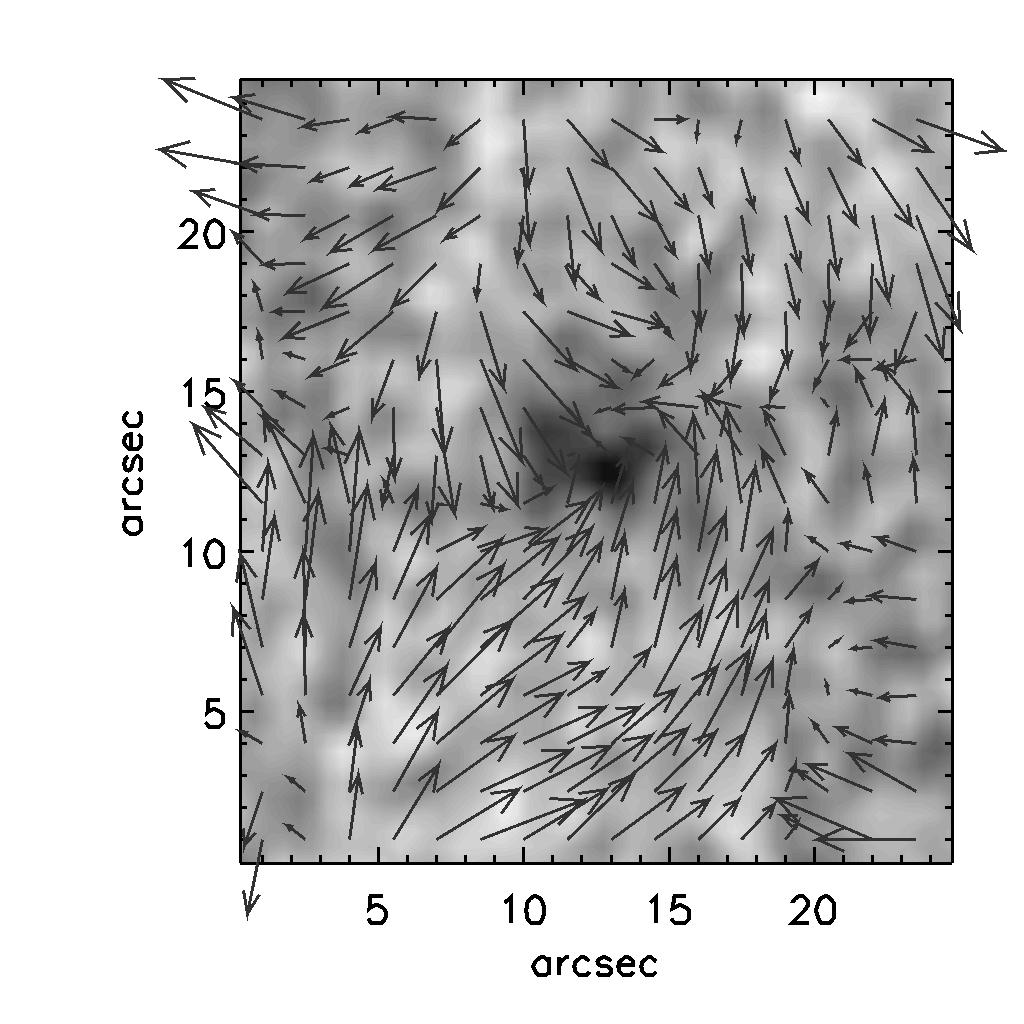}
   \includegraphics[width=0.35\textwidth,clip=]{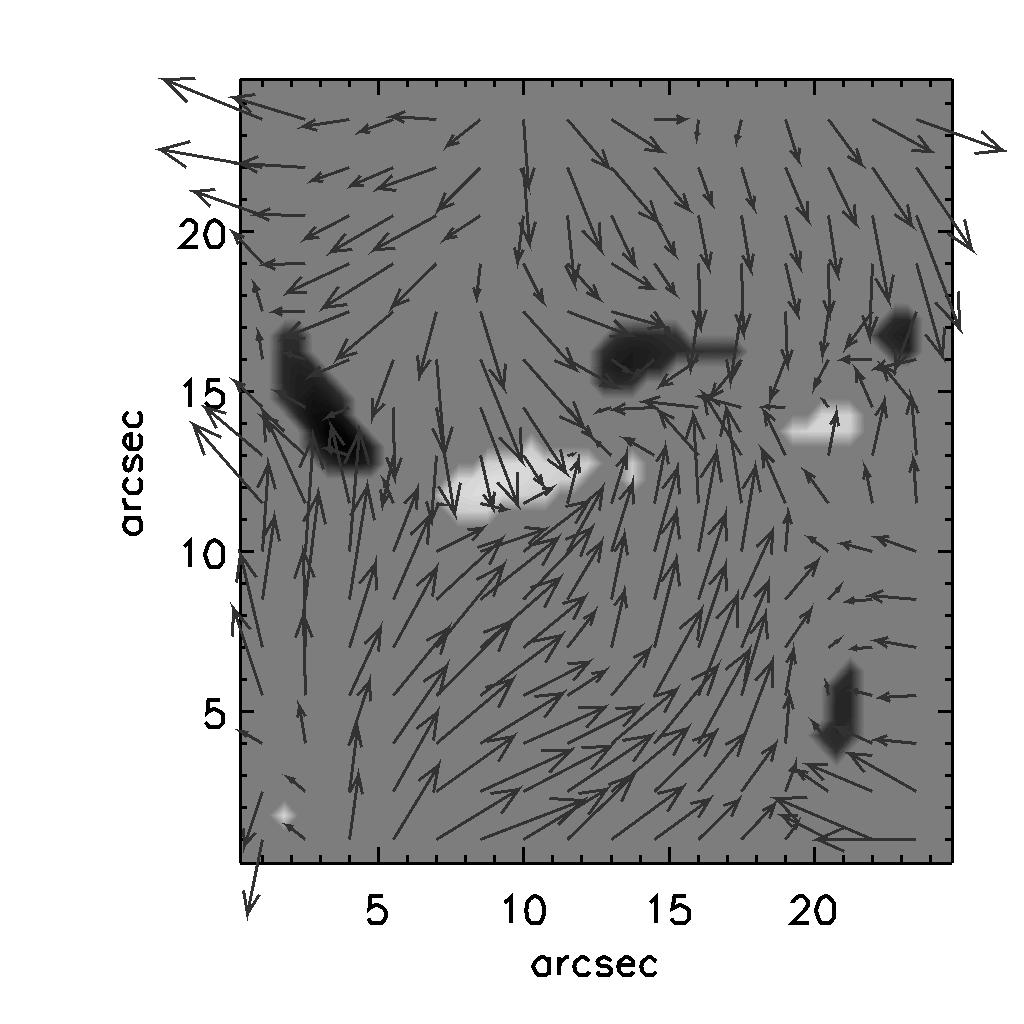}}
   
 \caption{
      {\bf Left:} Mean Doppler downflow number~8 between time 5~h and 10.7~h overlapped with the mean horizontal velocities.
      {\bf Right:} Vorticity characterised by the $\normG > 0.5$ parameter (downflow~8). White and black represent counter-clockwise $\Gamma_1$  and clockwise velocities, respectively.
  }
 \label{down8_gam}
\end{figure*}

\cite{Req2018} analysed a long-lived and large vortex (having a size of about 5~Mm) using high spatial (0.2\arcsec{}) observations from the Hinode satellite. In the literature, the terms 'vortex' and 'swirl' are both employed to identify rotating structures in the flows. A large review of the vortex (or swirls) detection is given in \citet{Maria2020}. Photospheric vortex flows are usually correlated with a network of magnetic elements at supergranular vertices \citep{Req2018}.

\subsection{Vortex detection}

Here, medium spatial resolution observations are used over a larger field of view to detect a potential vortex (or swirls). From the horizontal velocity field, we applied the vortex detection to identify 'swirls'  as described in  \citet{giag2017, Souze2018, liu2019}, with the two dimensionless parameters $\Gamma_1 \left(P \right)$ and $\Gamma_2 \left(P \right)$ at the target point $P$:
\begin{align}
    \Gamma_1 \left( \mathrm{P}\right) &= \frac{1}{N} \sum \limits_{M \in S} \frac{\vec{n}_{\mathrm{PM}} \times \vec{v}_{\mathrm{M}}}{\norm{\vec{v}_{\mathrm{M}}}},\\
    \Gamma_2 \left(\mathrm{P} \right) &= \frac{1}{N} \sum \limits_{M \in S} \frac{\vec{n}_{\mathrm{PM}} \times \left( \vec{v}_{\mathrm{M}} - \vec{v}_{\mathrm{P}} \right)}{\norm{\vec{v}_{\mathrm{M}} - \vec{v}_{\mathrm{P}}}},
\end{align}
where $\vec{v}_{\mathrm{M}}$ and $\vec{v}_{\mathrm{P}}$ are the velocity vectors at the points $\mathrm{M}$ and $\mathrm{P}$, $S$  is the two dimensional region with the size $N$ pixels surrounding the target point $\mathrm{P}$,  $\mathrm{M}$ is the point within the region $S$, $n_{\mathrm{PM}}$ is the normal vector pointing from $\mathrm{P}$ to $\mathrm{M}$, and $\times$ stands for the vector product \citep{liu2019}.

Here for convenience, we use only the term vortex and the parameter $\Gamma_1$ to detect the core of a vortex.  More sophisticated vortex detection have been recently developed (LCS approach; see \citealt{Souze2018} or \citealt{Chian2019}) but our small sample allows us to control the vortex detection by eye. For an ideal vortex, $\Gamma_1 \left(P \right)$ is maximum (value is equal to 1) at the centre of the vortex and decreases towards zero outside of it. Hence, it allows us to determine both the vortex location and the radius. Figure~\ref{Gam_29nov} shows the 13 persistent downflows relative to the measured $\Gamma_1$ parameter computed with the 24h mean velocity field. In that figure only a few of the 13 downflows correspond to the detected vortex (here via $\Gamma_1 > 0.5$). However, a detailed inspection reveals a vorticity, at least, in six of them (46\%) (downflows 1, 2, 4, 5, 8 and 13) during a portion of their lifetimes, including 9.9~h--17.3~h, 0.9~h--7.1~h, 10.9~h--18.1~h, 10.7~h--19.7~h,  and 10.9~h--16.3~h, respectively. Examples of downflows number~5 and 8 are given in the figure.

Figure~\ref{down5_gam} shows the mean Doppler and $\normG > 0.4$ parameter for the two periods of time relative to the beginning of the sequence: 11.4~h--15.2~h and 15.4~h--19.7~h of  downflow number~5. The temporal evolution of the flows indicates, very close to the downdraft, a direction change in the vortex rotation. Therefore, the vortex direction seems to be very sensitive to the evolution of local large-amplitude flows and can be reversed  as in our example. Another vortex is presented in Figure~\ref{down8_gam}, where we observe a clear link between $\normG > 0.23$ and persistent downflow number~8. One important point is that we never detect a vortex before the appearance of the magnetic field in the downflow, when it is possible to observe that in a few cases. We do not know if this is due to the 1\arcsec{} of spatial resolution and 45s temporal step.

\subsection{Requerey's vortex and long-lived downflows }

\begin{figure}
    \includegraphics[width=0.5\textwidth,clip=]{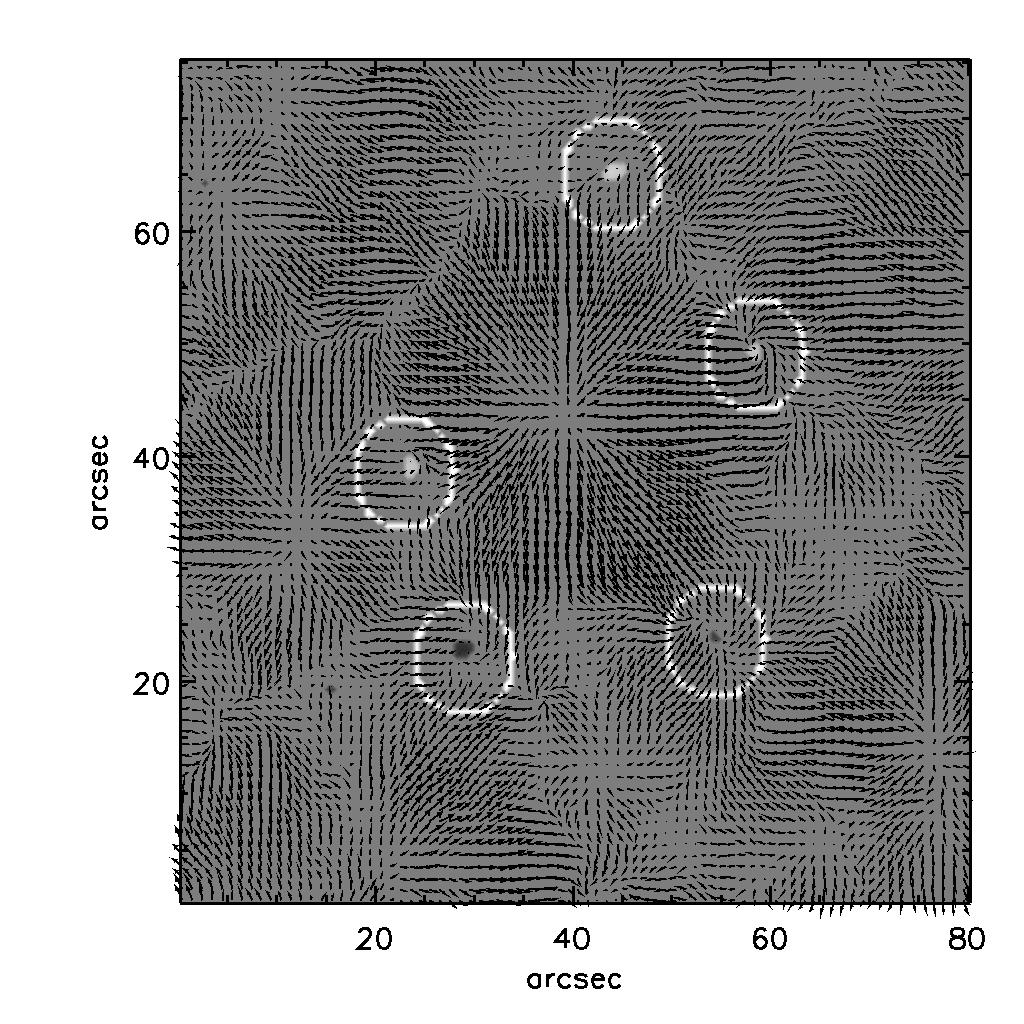}
    \caption{$\Gamma_1$ 'vortex' (in circles) overlapped by horizontal velocity field. }
    \label{Gam_req}
\end{figure}

\begin{figure}
    \centerline{\includegraphics[width=0.5\textwidth,clip=]{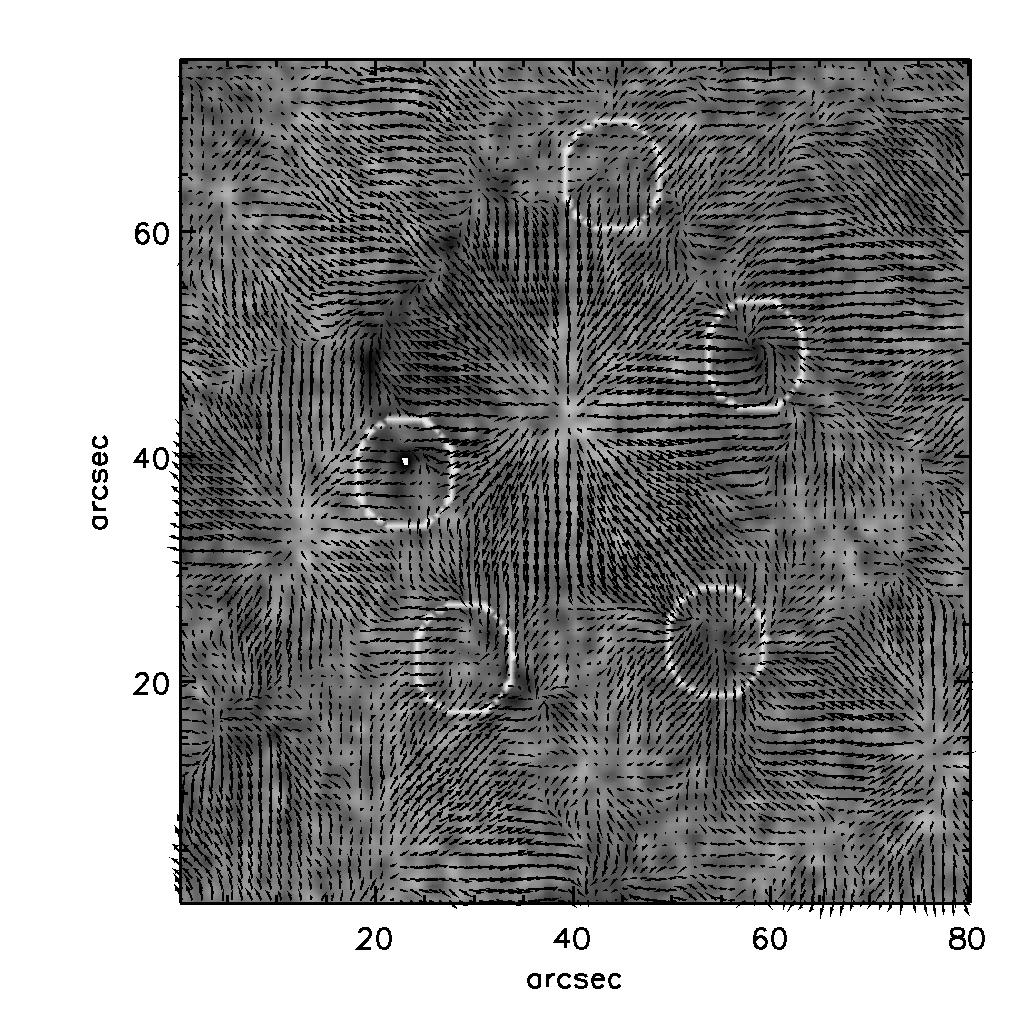}}
    \caption{Mean Dopplergram overlapped by horizontal velocity field and the location of the $\Gamma_1$ 'vortex' (in circles).}
    \label{Dop_req}
\end{figure}

\begin{figure}
    \centerline{\includegraphics[width=0.5\textwidth,clip=]{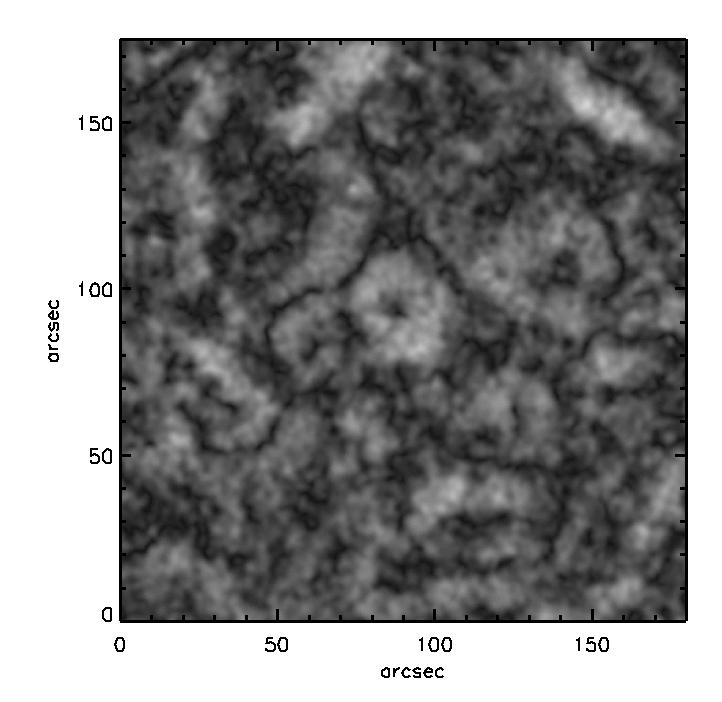}}
    \caption{Requerey's vortex in a large field of view of the module of the mean horizontal velocities (24-h mean)  }
    \label{mod_vel_req}
\end{figure}

\begin{figure}
    \centerline{
    \includegraphics[width=0.3\textwidth,clip=]{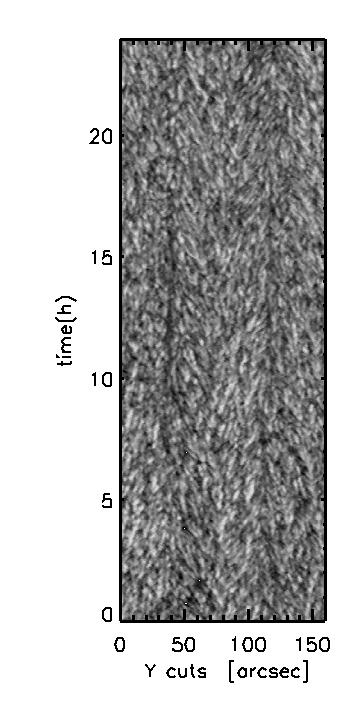}
   \includegraphics[width=0.3\textwidth,clip=]{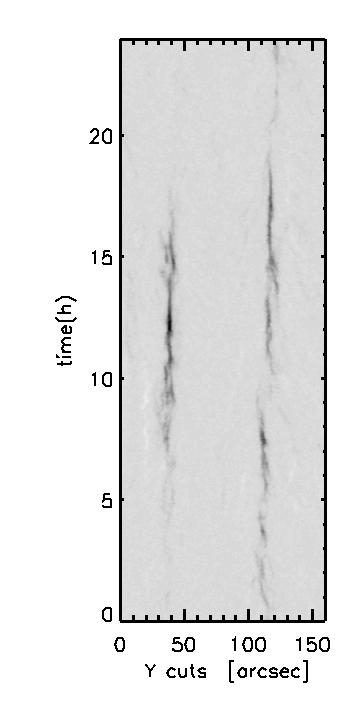}}

    \caption{Persisting downflows during 24~h. $\left(y, t\right)$ cuts through the Dopplergram (left) and \Bpar{} (right). Requerey's vortex corresponds to the long-lasting downflow and \Bpar{} is located around $y \approx 120\arcsec{}$. }
    \label{cuts_vel_req}
\end{figure}

To validate our vortex detection via the parameters $\Gamma_1$ and $\Gamma_2$, we used the SDO observation used in the analysis of  \cite{Req2018, Souze2018, Chian2019}. Luckily, SDO observations started just before Requerey's observations were made with the Hinode satellite. We remind the readers that   \cite{Req2018, Souze2018, Chian2019} observed and analysed a nice long-lasting vortex. The advantage of the SDO observations is the access to various kinds of context observations, on the same day, to a larger field of view. Despite the spatial resolution of SDO being lower than that of Hinode, we detected  the vortex described in \cite{Req2018} without any problem. The locations of  $\Gamma_1$, vortex detection overlapped by a horizontal velocity field are plotted in Fig.~\ref{Gam_req}. The Requerey's vortex is well detected by the gamma-method and is clearly visible in the plotted horizontal velocities (at the coordinates $\left(62\arcsec \times 49\arcsec\right)$. Figure~\ref{Dop_req} shows the mean Doppler overlapped by horizontal velocity field and the location of the $\Gamma_1$ corresponding to downflows around the supergranule. 

In our observations, five larger-amplitude regions in the $\Gamma_1$ parameter are seen in total (see Fig.~\ref{Gam_req} and \ref{Dop_req} ) and only one of them corresponds to the vortex studied by Requerey. The others have vorticity but also a non-negligible shear component which contributes to the final amplitude of $\Gamma_1$ and hence does not allow for one to call them a vortex. However, \cite{Chian2019, Chian2020} with higher spatial resolution, reported on the same field of view, more persistent objective vortices with shorter lifetimes corresponding to the gap regions of 'Lagrangian chaotic saddles'.

In the larger field of view, the horizontal velocity module (Figure~\ref{mod_vel_req}) shows, in the central part, the roundish supergranule corresponding to the supergranule shown in  Figure~\ref{Gam_req}, where  Requerey's vortex is observed. This supergranule is a particularly symmetrical, well-formed supergranule with large amplitude horizontal outflows. The large horizontal velocity amplitude of that supergranule and the combination of the velocities from the large supergranules on its right location on the  eastern edge, produce the large Requerey's vortex. The temporal and spatial coincidence is probably the reason for producing this  large and long-lasting vortex. The other large amplitude downflows around this supergranule do not show identical properties.

\begin{figure}
    \centerline{\includegraphics[width=0.5\textwidth,clip=]{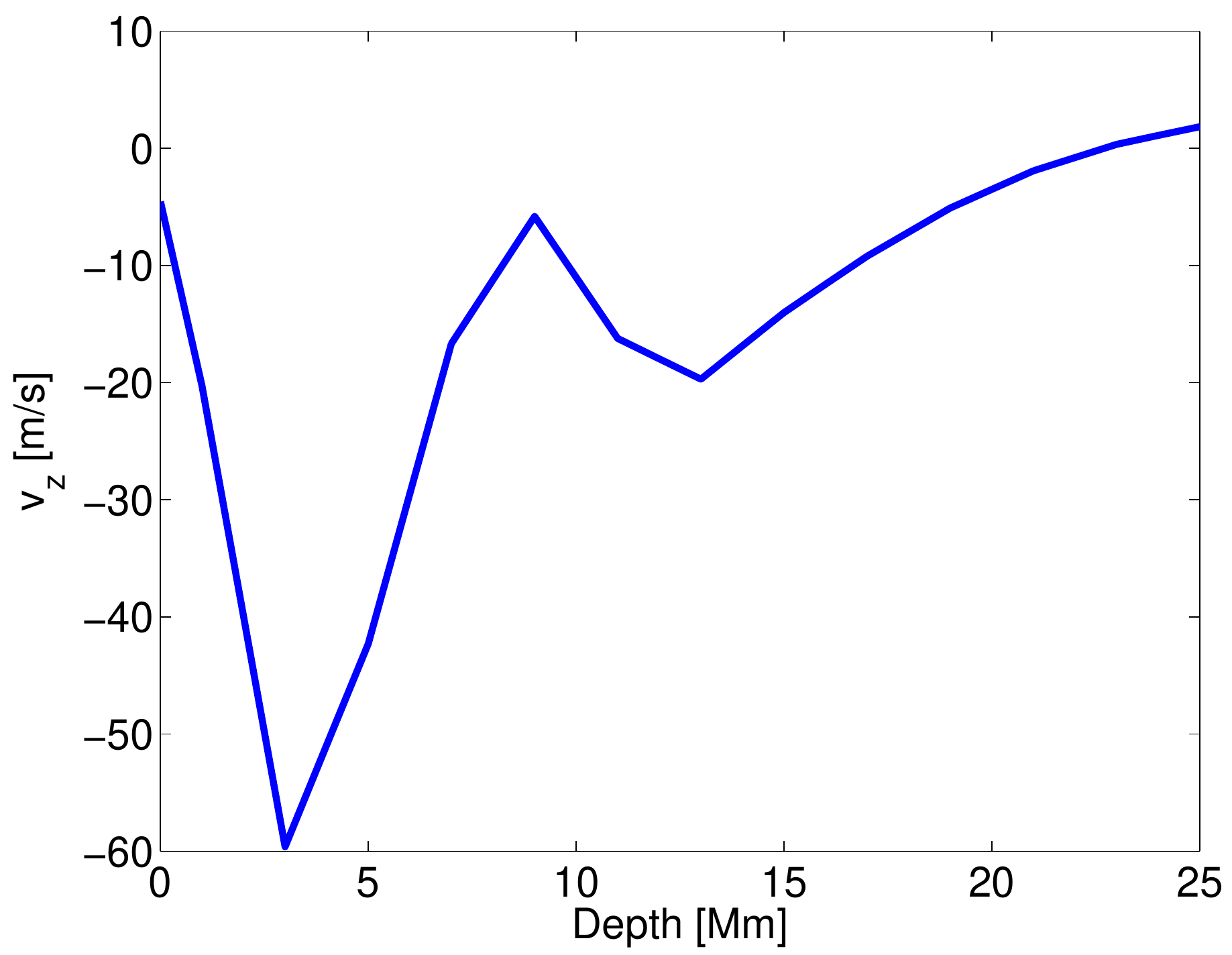}}
    \caption{Depth profile of the 24h averaged vertical velocity in the middle of  Requerey's vortex. }
    \label{Vz_prof}
\end{figure}

\begin{figure}
    \centerline{\includegraphics[width=0.5\textwidth,clip=]{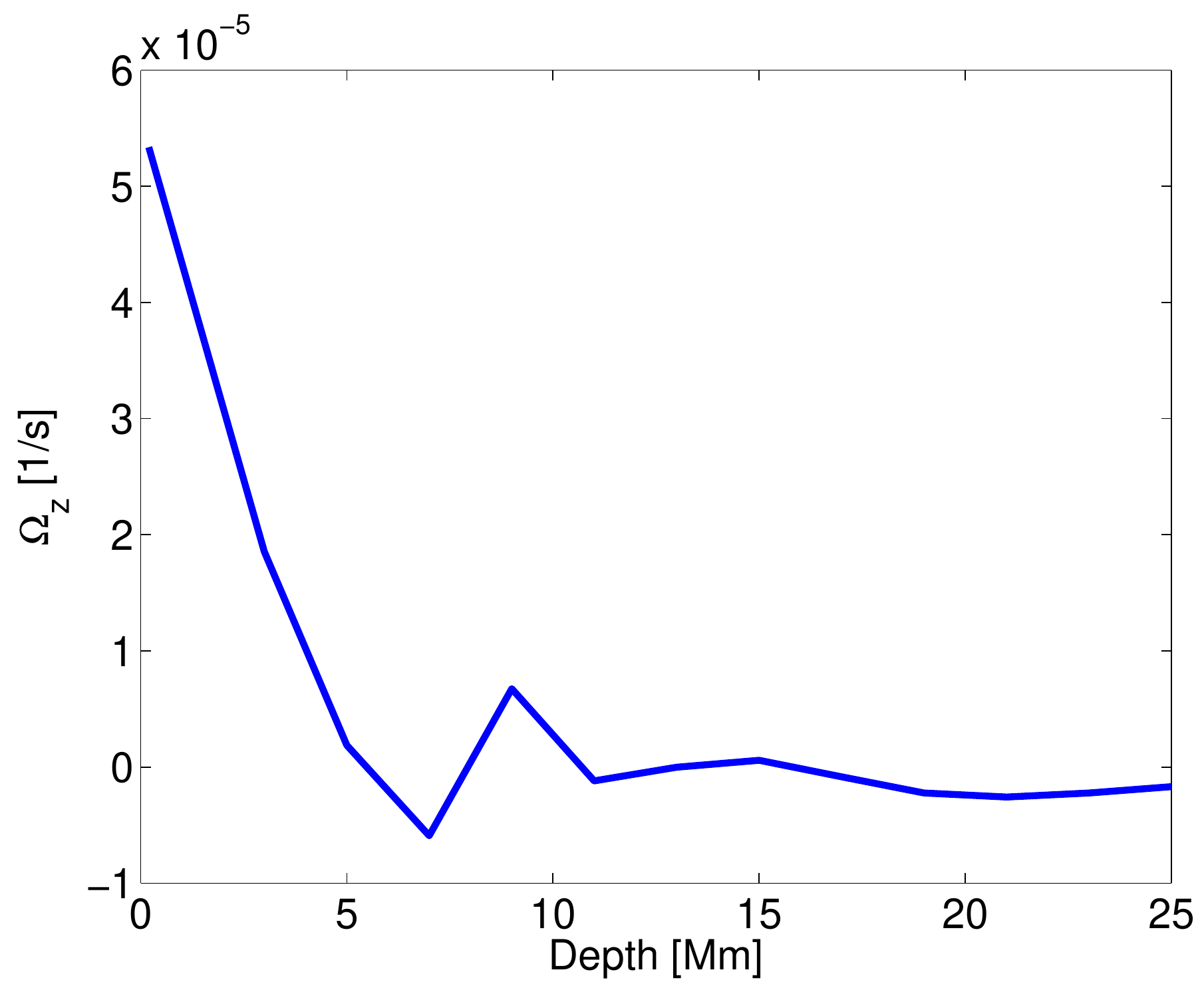}}
    \caption{Depth profile of the vertical component of the flow field vorticity in the middle of Requerey's vortex.}
    \label{Vort_prof}
\end{figure}

Helioseismic inversion allows us to study the general properties of the large Requerey's vortex in depth. In the map of the inverted component, nothing extraordinary is seen at the location of this vortex, likely because the spatial resolution of the flow maps is again coarser than the extent of the vortex. The plot of the vertical profile of the vertical velocity, on the other hand, clearly shows that the downflow extends deep into depth of 20~Mm and more (see Fig.~\ref{Vz_prof}). The plotted profile is similar to the profile of the strongest downflows recorded in the studied field of view on 29 November 2018 (compare to the dash-dotted line in Fig.~\ref{mean_vz_cuts}).

In the maps of the vertical vorticity computed from the horizontal flow components as
\begin{equation}
    \Omega_z=\frac{\partial v_x}{\partial y}-\frac{\partial v_y}{\partial x},
\end{equation}
 Requerey's vortex is clearly visible. The depth profile of $\Omega_z$ shows that the vorticity keeps its integrity to the depth of about 5~Mm (see Fig.~\ref{Vort_prof}), where it seems to start to oscillate around zero and vanishes at the depth of about 11~Mm. We do not think that the peak at the depth of 9~Mm that is visible in both vertical velocity and vertical vorticity profiles is particularly real. The formal uncertainty of the vertical velocity at this depth is about 10~\mps{} and that of the vertical velocity vorticity about $10^{-5}$ s$^{-1}$, which corresponds to the magnitude of these 'peaks'. The comparison of Requerey's vortex and the downflows in the field of view on 29 November 2018 indicates that the long-living vortex is linked to strong downflows.

\section{Simulation link between magnetic field and persistent downflows}
\label{sec:link_downflows_B}

We used two different datasets  based on the 3D MHD numerical simulation. These runs apply to the purely quiet Sun \citep{SN98}.

The first dataset provides the velocity vector $\vec{v} \left(x, y, t \right)$ and the magnetic field vector $\vec{B} \left(x, y, t \right)$ over the solar surface $\left( z = 0\right)$ for 27 hours (spatial resolution: 96~km/pixel; temporal resolution: 60~s; field of view (FOV): $96~\mathrm{Mm}\times96~\mathrm{Mm}$). The corresponding resolution is 0.13\arcsec{}  on the Sun, as the computational pixel size is 48~km, which is much better than observations. We used  the $v_x$, $v_y$, $v_z$ as well as $B_z$ quantities. The $z$ axis is positive above the surface.
The second dataset, which is much shorter, gives only the vertical velocity $v_z \left(x,y,z,t \right)$ during 4 hours from the surface $\left( z = 0\right)$ to $z = - 20$~Mm, with the same horizontal FOV and time step as the first dataset.

In order to eliminate 5-minute oscillations, the vertical velocity field $v_z$ was filtered in the \kw{} diagram low frequencies such that $\omega$ < $c$ $k$ were kept and other frequencies were eliminated ($c = 6$~k\mps{}). Then,  we averaged velocities and magnetic fields over time intervals 12 and 24~hours for the first sequence close to the supergranulation lifetime and 4~hours for the second sequence in order to keep only long-lasting features.

\begin{figure*}
    \centerline{\includegraphics[width=1.0\textwidth,clip=]{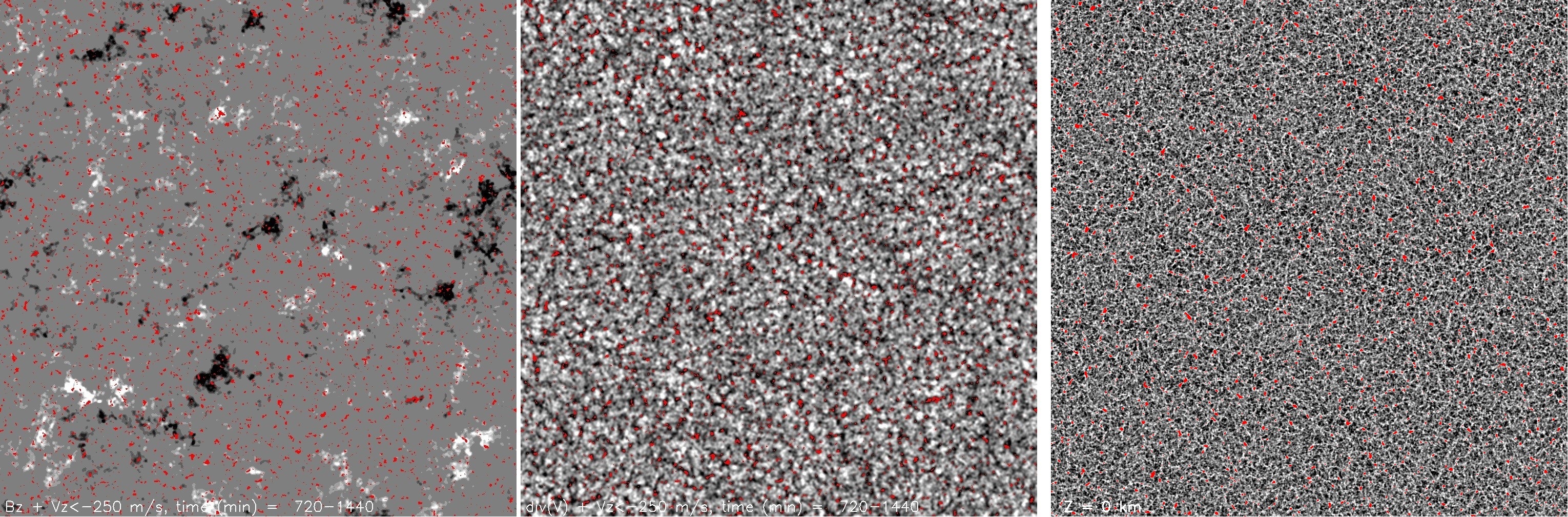}}
    \caption{$B_z$, $\vec{\nabla}_h \cdot \vec{v}_h$, $\vec{\nabla} v_z$ at the surface (FOV $96\times96$~Mm$^2$). Long-lasting downward velocities are displayed as red points using the threshold of $-0.25$~k\mps{}.
    {\bf Left:} $B_z$ (grey levels), averaged over 12~hours.
    {\bf Middle:} $\vec{\nabla}_h \cdot \vec{v}_h$  averaged over 12~hours. Bright and dark  correspond to diverging and converging flows, respectively.
    {\bf Right:} $\mathrm{d}v_z/\mathrm{d}z$  averaged over 4~hours. Bright and dark correspond to upward and downward vertical gradient, respectively. 
    }
    \label{BV}
\end{figure*}

\begin{figure*}
    \centerline{\includegraphics[width=1.0\textwidth,clip=]{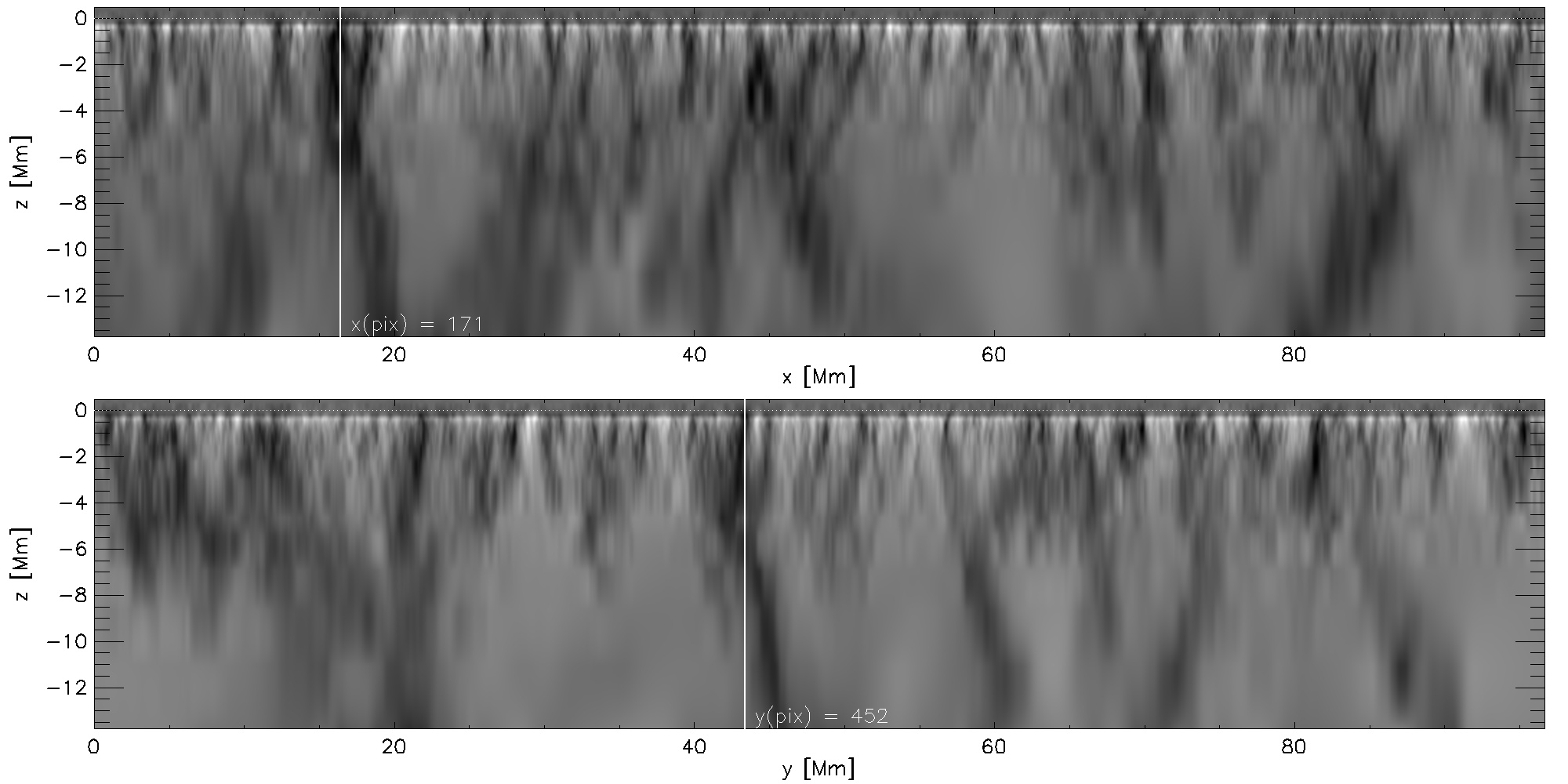}}
    \caption{Cross sections of the vertical velocity $v_z$ (in $x \times z$ and $y \times z$ planes) at pixel $\left(x = 171~\mathrm{Mm}, y = 452~\mathrm{Mm}\right)$ indicated by the vertical lines, where a typical downflow occurs. The surface is at $z = 0$~Mm. Bright and dark correspond to upflows and downflows, respectively.}
    \label{Vz_cuts_simul}
\end{figure*}

Figure~\ref{BV} displays 12-hour averages (first dataset) of the vertical magnetic field $B_z$ and  the divergence of the horizontal velocities $v_h = \left(v_x, v_y \right)$. The long-duration downflows (red points) are spatially much more concentrated and also much more numerous than in observations. They occur almost everywhere, even in non-magnetic areas. Such points are systematically characterised by 
$\vec{\nabla}_h \cdot \vec{v}_h < 0$ (converging flows) with the mean value of $-1.8\times 10^{-3}$~s$^{-1}$. The second dataset shows that these points are also characterised by $\mathrm{d}v_z/\mathrm{d}z > 0$ at z = 0, which means that the amplitude of $v_z$ decreases with height in these points at the surface. The mean value is $1.9\times 10^{-3}$~s$^{-1}$, so that we are close to $\vec{\nabla} \cdot \vec{v} = 0$ at the surface. Therefore, the fluid exhibits an uncompressible behaviour in this layer.

Figure~\ref{Vz_cuts_simul} shows two typical cross sections of the vertical flows  based on the second dataset, averaged over 4 hours. Some downflows extend from the surface down to $z = -13$~Mm. However, most downflows are not deeper than $z = -4$~Mm. We also point out that the continuous downflow regions do not extend purely vertically to a certain depth, but they depict a lateral displacement between the depths. This behaviour is very similar to what we found in the helioseismic inversions (cf. Fig.~\ref{Vz_cuts_examp}).

We conclude that the numerical simulation predicts the existence of spatially concentrated lasting downflows in the quiet Sun with a typical size of about 0.25\arcsec{}. However, the spatial resolution of observations used in this paper does not allow us to confirm simulation results. In the simulations, we did not find evidence of a vortex comparable in size to the observed vortex on the surface of the Sun.

\section{Results and conclusion}
\label{sec:results}

The properties of solar plasma below the surface is mostly inferred from the observations of solar surface oscillations. 
The formation of vigorous downflowing plumes and the collective interaction of these \citep{R03a, Cos2016} also provides the opportunity, combined with local helioseismology, to probe the first megameters of the solar interior. Lot of observations of vortices (vortex)  associated with downflows are described in the literature \citep{Attie2009, Duvall2010, Bonet2010, Vargas2011, Vargas2015, Attie2016, Req2017, Req2018}, generally studied on short sequences. Three of them are observed on long temporal sequences from 4~h to 24~h \citep{Attie2009, Duvall2010, Attie2016, Req2018}. Taking advantage of the long temporal and homogeneous SDO/HMI observations (24~h here) of the Dopplergrams and magnetograms, we investigate the properties of the downflows, particularly the long-lasting one called persistent downflows here, and their links to a concentrated magnetic field and finally the link with the corona observations (AIA 193~\AA). 

In the 3D data cube $\left(x,y,t\right)$ of the 24-hour Doppler sequence (29 November 2018), at the disc centre, we detect 13 persistent downflows giving a rate of occurrence of $2\times 10^{-4}$ cases per~Mm$^{2}$ and 24~h. This rate is lower than  those found by \citet{Req2017} during sequences of 32.0 and 22.7 minutes of $6.7\times 10^{-2}$ per Mm$^{2}$ which was found with higher spatial resolution but on shorter time sequences. The lifetimes of our 13 persistent downflows were found between 3.5~h to and 20~h with a sizes  between 2 and 3\arcsec{} and speeds between $-0.25$ and  $-0.72$~k\mps. They are well located at the junction  of several supergranules as observed by  \cite{Attie2009, Attie2016, Chian2019, Chian2020}. In our examples, only 46\% of the persistent downflows were connected with an observed vortex, but not all the time and not for all the studied downflows.  At higher spatial resolution \citep{Chian2019, Chian2020}, more persistent objective vortices were found with shorter lifetimes corresponding to the gap regions of 'Lagrangian chaotic saddles'.

All the studied persistent downflows are always  associated with a magnetic field (up to 600 Gauss).  No significant downflows were observed before the appearance of the magnetic field as described in \cite{Vargas2015}. The advantage of their higher spatial resolution (0.15\arcsec{}) probably allows them to detect downflows before the magnetic field appearance and follow the enhancement of this magnetic field by the strong downflows. The detection of the vortex described by \cite{Req2018} with our SDO/HMI data gives confidence in our method of vortex detection. This very beautiful vortex, with our large field of view, seems to be the result of the composition of the larger-scale horizontal flows  where the combination of the flow phases, temporal and spatial, creates it.

We observe that this downflow is always associated with the presence of the magnetic field. The helioseismic inversion allows us to describe the persistent downflow properties in depth below the solar surface and, for instance, to compare the properties of the persistent downflows to the other (non-persistent) in the field of view. Persistent downflows seem to penetrate deeper, whereas the common downflows seem to reach zero vertical velocity already at the depth of around 7~Mm. The persistent downflows remain slightly negative on average all the way to the bottom of our region of interest, which is 25~Mm. 
 
In the high spatial resolution MHD simulation (0.13 \arcsec{}), long-duration downflows (4~h and 12~h) are spatially much  smaller and numerous than in observations where the resolution lies around 2 to 3\arcsec{}. We did not find evidence of a vortex comparable in size to observations. The spatial resolution of observations does not allow us to see small scale downflows, so that observations with more precise satellites (as the PHI on-board Solar Orbiter) or ground-based large telescopes are required.

From the space-borne SDO instruments, downflow evolution is studied at different wavelengths. We observe a persistent downflow to be correlated with the coronal-loop anchor in the photosphere, indicating a link between  downflows and the coronal activity. This observation is to be compared with the observed EUV cyclones over the quiet Sun, anchored in the rotating magnetic field  network magnetic fields, suggesting an effective way to heat the corona \cite{Zhang2011}. The link between the flow evolution at the junctions  of supergranular cells, giving mini-coronal-mass ejection and X-ray network flares, is described in \citet{Innes2009, Attie2016} . Hence, the persistent downflows must be investigated with a greater number of time sequences.In the same way, more  studies have to be performed on the link between persistent downflows and the coronal phenomena to confirm the link between them and describe the temporal dynamic behaviour of such structures. 

 The new non-linear methods (Lagrangian Coherent Structure and Lagrangian Chaotic Saddles) developed recently by \cite{Chian2019, Chian2020},
applied to the full Sun SDO/HMI observations, is probably the best way for future research in this domain to access to  large-events statistics. 
 
\begin{acknowledgements}
    This work was granted access to the HPC resources of CALMIP under the allocation 2011-[P1115]. Thanks to SDO/HMI, IRIS, and Hinode/SOT teams. M\v{S} and DK were supported by the Czech Science Foundation under the grant project 18-06319S. 

\end{acknowledgements}

\end{document}